\documentclass[a4paper,11pt]{article} 

\pdfoutput=1

\usepackage[x11names, svgnames, rgb]{xcolor}

\usepackage[T1]{fontenc}
\usepackage[english]{babel}
\usepackage{tikz}
\usepackage[compat=1.1.0]{tikz-feynman}
\usepackage{slashed}
\usepackage{tabularx}
\usepackage{wrapfig}
\usepackage{subfig}
\usepackage{subcaption}
\usepackage{afterpage}
\usepackage{physics}
\usepackage{comment}
\usepackage{emptypage}
\usepackage{listings}
\usepackage{pdfpages}
\usepackage{enumerate}
\usepackage{appendix}
\usepackage{bbold}
\usepackage{bm}
\usepackage{mathrsfs}
\usepackage{setspace}
\usepackage{pifont}
\usepackage{xfrac}
\usepackage{cancel}
\usepackage{mathtools}
\usepackage{multirow}
\usepackage{booktabs}
\usepackage{makecell}
\usepackage[export]{adjustbox}
\usepackage{lmodern}
\usepackage{soul}

\usetikzlibrary{snakes,arrows,shapes}
\usepackage{jheppub} 

\newcommand{\agl}[2]{\langle#1 #2 \rangle}
\newcommand{\sqr}[2]{\lbrack #1 #2 \rbrack}

\newcommand{\tildee}[1]{\widetilde{#1}}

\DeclareMathOperator*{\dime}{\text{dim}}
\makeatletter\g@addto@macro\bfseries{\boldmath}\makeatother

\definecolor{nicered}{rgb}{0.7,0.1,0.1}
\definecolor{nicegreen}{rgb}{0.1,0.5,0.1}
\definecolor{violet}{rgb}{0.7,0.3,0.3}
\hypersetup{colorlinks,citecolor= nicegreen,linkcolor= nicered}

\title{Anomalous Dimension of a General Effective Gauge Theory II: Fermionic Sector}

\author[a]{Jason Aebischer,}
\author[b,c]{Luigi C.~Bresciani,}
\author[b,c]{Nud\v zeim Selimovi\'c}
\affiliation[a]{PSI Center for Neutron and Muon Sciences, 5232 Villigen PSI, Switzerland}
\affiliation[b]{Dipartimento di Fisica e Astronomia ``G.~Galilei'', Università degli Studi di Padova,\\Via F.~Marzolo 8,
I-35131 Padova, Italy}
\affiliation[c]{Istituto Nazionale di Fisica Nucleare, Sezione di Padova,\\Via F.~Marzolo 8,
I-35131 Padova, Italy}

\emailAdd{jason.aebischer@psi.ch}
\emailAdd{luigicarlo.bresciani@phd.unipd.it}
\emailAdd{nudzeim.selimovic@pd.infn.it}

\abstract{The complete set of one-loop anomalous dimensions for general Effective Field Theories (EFTs) is derived using on-shell methods. Combined with previous findings for the bosonic sector, the obtained results conclude the computation of the complete set of leading order Renormalization Group Equations (RGEs) in arbitrary gauge EFTs containing scalar and fermion fields. Renormalization effects are consistently taken into account at the order $1/\Lambda^2$ in the new physics scale $\Lambda$ for all renormalizable and non-renormalizable couplings. The obtained template RGEs include operator mixing across different dimensions and are valid for arbitrary gauge groups. 
}
\allowdisplaybreaks

\begin{document}
\maketitle
\pagestyle{myplain}

\section{Introduction}
\label{sec:intro}
Effective Field Theories (EFTs) are a powerful framework for studying phenomena that lie beyond the Standard Model (SM) of particle physics. They enable a model-independent analysis of New Physics (NP) effects, relying only on a handful of assumptions. One core premise of EFTs is that they are only valid up to a given cut-off scale $\Lambda$. As a consequence of the decoupling theorem the effective couplings of EFTs are suppressed by inverse powers of the NP scale $\Lambda$. This introduces a natural power counting parameter $\epsilon:=f/\Lambda$, where $f$ stands for small momenta or mass scales in the EFT. Predictions can therefore be made in a consistent way within the EFT framework up to a given order in $\epsilon$, neglecting higher-order contributions. In this article we will work up to order $\epsilon^2$ and therefore only retain terms in the calculation that are at most $\mathcal{O}(1/\Lambda^2)$. 

Another feature of EFTs, as in any Quantum Field Theory, is that the parameters depend on the renormalization scale. This scale dependence is governed by the Renormalization Group Equations (RGEs) of the theory, which are related to the additional divergences introduced by the EFT operators. Consequently, the RGEs depend on the field content and the imposed gauge symmetry and therefore need to be derived anew for each individual EFT. It is however possible to adopt a general EFT framework that encompasses a general field content and gauge group. In this setup the RGEs can be computed once and for all, yielding general template RGEs. Specific RGEs for a given EFT can then be deduced from the template RGEs by algebraic and group theory manipulations, without the need to do any loop computations. 

For the dimension-four interactions several template RGEs exist in the literature. Some early work on this topic at the two-loop level can be found for instance in \cite{Jack:1984vj,Machacek:1983tz,Machacek:1983fi,Machacek:1984zw,Luo:2002ti,Schienbein:2018fsw}. The two-loop renormalization of vacuum expectation values is given in \cite{Sperling:2013eva,Sperling:2013xqa}. The three-loop beta function for gauge and Yukawa interactions were derived in \cite{Pickering:2001aq,Poole:2019kcm,Mihaila:2012pz} and \cite{Davies:2021mnc,Poole:2019txl}, respectively. For models containing scalars and fermions the three-loop gaugeless RGEs were derived in \cite{Jack:2023zjt,Steudtner:2021fzs} and the quartic scalar coupling was renormalized at three loops in \cite{Steudtner:2024teg}. The four-loop generic gauge beta function was computed in \cite{Bednyakov:2021qxa} and the running of the renormalizable couplings of scalar theories is known up to six loops \cite{Bednyakov:2021ojn,Bednyakov:2025sri}.

Concerning higher-dimensional operators, in \cite{Aebischer:2025zxg} (see also \cite{Misiak:2025xzq,Fonseca:2025zjb}), the complete one-loop running for bosonic theories\footnote{Recently, even the two-loop renormalization for general bosonic effective theories up to dimension six was computed \cite{Guedes:2025sax}.} up to mass dimension six was derived using the on-shell methods developed in~\cite{Caron-Huot:2016cwu}.\footnote{See also \cite{Chavda:2025aqm} for a recently proposed connection between unitarity and renormalization flow.} The key point behind the framework in~\cite{Caron-Huot:2016cwu} is that it targets the renormalization-scale dependence appearing through logarithmic terms in scattering amplitudes. Since these logarithms involve kinematic quantities with non-trivial mass dimensions, their scale dependence must be balanced by that of the Wilson coefficients, allowing anomalous dimensions to be extracted directly from the cut amplitudes. This approach has already yielded results in the context of the Standard Model Effective Field Theory (SMEFT)~\cite{Ma:2019gtx,Jiang:2020mhe,AccettulliHuber:2021uoa,EliasMiro:2020tdv}, as well as in EFTs containing axion-like particles~\cite{Bresciani:2024shu}, while the method has been generalized to include the mixing of operators with different dimensions in~\cite{Bresciani:2023jsu}.

A key strength of the on-shell formalism lies in its conceptual and practical simplicity. First, calculations are performed exclusively with physical external states, thereby avoiding any mixing with unphysical operators. Second, gauge invariance is preserved manifestly throughout the computation. Finally, powerful selection rules~\cite{Cheung:2015aba,Bern:2019wie,Jiang:2020rwz,Chala:2023xjy} governing scattering amplitudes sharply restrict the allowed operator mixings, leading to a substantial reduction in computational complexity. These features make the on-shell approach particularly effective for determining renormalization group equations in general EFTs~\cite{DeAngelis:2022qco}, yielding results that are directly usable without invoking field redefinitions or additional simplifications associated with gauge redundancies~\cite{Baratella:2020lzz}.
Beyond one loop, the unitarity-based extraction of anomalous dimensions has also proven to be a viable and powerful tool. Two-loop computations performed within this framework represent the current frontier in EFT renormalization, with a number of results already available in the literature~\cite{Panico:2018hal,EliasMiro:2021jgu,Bern:2020ikv}.

In this article, we generalize the work in \cite{Aebischer:2025zxg} by including the RGE effects of fermionic operators at the one-loop level. For this purpose, we adopt the most general physical basis of operators up to mass dimension six that contains an arbitrary number of scalar, fermion, and vector fields, which are invariant under an arbitrary gauge group. Such a physical basis of operators has already been constructed in~\cite{Aebischer:2025zxg} by identifying the independent kinematic structures associated with the contact amplitudes in the EFT~\cite{Shadmi:2018xan,Cheung:2016drk,Falkowski:2019zdo,Durieux:2019siw,Li:2022tec,DeAngelis:2022qco,Durieux:2019eor}.

In this setup, the RGEs are computed for all Lagrangian parameters, up to $\mathcal{O}(1/\Lambda^2)$ in the NP scale $\Lambda$. The results of this article conclude the program of general one-loop RGEs for effective field theories up to mass dimension six. For any conceivable EFT containing scalars, fermions and vectors that is invariant under an arbitrary gauge group the RGEs can be constructed using the template RGEs derived here and in \cite{Aebischer:2025zxg}. 

The rest of the article is structured as follows: In Sec.~\ref{sec:EFT} we define the complete Lagrangian used for the computation of the RGEs, together with the used notation. Sec.~\ref{sec:method} contains a brief review of the employed on-shell method to obtain the RGEs. In Sec.~\ref{sec:results} we present the obtained results, and conclude in Sec.~\ref{sec:concl}. Conventions and definitions are collected in App.~\ref{app:conventions}.

\section{General effective gauge theory}
\label{sec:EFT}

In this section, we introduce the general Lagrangian used to compute the template RGEs. It contains the renormalizable part, as well as the dimension-five and -six terms and reads
\begin{equation}\label{eq:fullEFT}
    \mathcal{L}_{\text{EFT}} = \mathcal{L}^{(4)}+\mathcal{L}^{(5)}+\mathcal{L}^{(6)}\,.
\end{equation}
The individual parts of the full Lagrangian are specified in subsequent subsections, specifically in Eqs.~\eqref{eq:L4}, \eqref{eq:L5}, and \eqref{eq:L6}, respectively. The full classification of Poincar\'e and gauge invariant operators up to dimension six can be found in \cite{Aebischer:2025zxg}.

\subsection{Renormalizable interactions}

We start by introducing the most general Lagrangian up to mass dimension four that contains an arbitrary number of real scalar fields $\phi_a$, gauge fields with field strengths $F_{\mu\nu}^{A_\alpha}$, and left-handed Weyl spinors $\psi_i$:\footnote{For the two-component notation we follow the conventions in \cite{Dreiner:2008tw}.}
\begin{align}\label{eq:L4}
\mathcal{L}^{(4)} &=  \mathcal{L}^{(4)}_{\text{bos}} + \mathcal{L}^{(4)}_{\text{fer}}\,,\\
    \mathcal{L}^{(4)}_{\text{bos}} &= -\frac{1}{4}\sum_{\alpha=1}^{N_G}\sum_{\beta=1}^{N_G}\xi^{A_\alpha B_\beta} F^{A_\alpha}_{\mu\nu}F^{{B_\beta}\,\mu\nu}+ \sum_{\alpha=1}^{N_G}\sum_{\beta=1}^{N_G} \tildee \xi^{A_\alpha B_\beta} F^{A_\alpha}_{\mu\nu}\tildee F^{{B_\beta}\,\mu\nu}  + \frac{1}{2} \eta_{ab} (D_\mu\phi)_{a}(D^\mu\phi)_b\nonumber\\
    &\quad -\Lambda-t_a \phi_a-\frac{m^2_{ab}}{2!}\phi_{a}\phi_{b}-\frac{h_{abc}}{3!}\phi_a \phi_b\phi_c
    -\frac{\lambda_{abcd}}{4!}\phi_{a}\phi_{b}\phi_{c}\phi_{d}\,,\\
    \mathcal{L}^{(4)}_{\text{fer}} &= i\, \kappa_{ij}\,  \psi^\dagger_i \overline{\sigma}^\mu (D_\mu \psi)_j - \frac{1}{2!}\left (m_{ij} \psi_i \psi_j + m_{ij}^* \psi_i^\dagger \psi_j^\dagger \right) - \frac{1}{2!}\left (y_{ija} \psi_i \psi_j + y_{ija}^* \psi_i^\dagger \psi_j^\dagger \right)\phi_a  \,.
\end{align}
Here, $\overline \sigma^\mu = (\mathbb{1}_{2\times 2}, -\vec \sigma)$, where $\vec \sigma$ are the Pauli matrices.
In addition to all kinetic and mass terms for scalars and fermions, the Lagrangian includes topological terms, the vacuum energy $\Lambda$, the tadpole interaction, triple and quartic scalar interactions, and Yukawa terms. In particular, the real parameters $m^2_{ab}$, $h_{abc}$, and $\lambda_{abcd}$ are totally symmetric, and the same is true for the complex fermion mass matrix $m_{ij}$ and the Yukawa matrices $y_{ija}$.

The gauge group $G$ is assumed to be a direct product of an arbitrary number of simple and compact gauge groups $G_\alpha$
\begin{equation}
    G = \prod_{\alpha=1}^{N_G} G_\alpha\,.
    \label{eq:full_gauge_group}
\end{equation}
In the following, the adjoint indices of the gauge groups $G_\alpha$ are denoted by $A_\alpha,B_\alpha,\dotsc \in \{1,\dotsc,\dime G_\alpha\}$,
while scalars and fermions are taken to transform in (possibly reducible) representations $S_\alpha$ and $F_\alpha$, respectively. 
The corresponding gauge indices are
$a_{\alpha},b_{\alpha},\dotsc \in \{1,\dotsc,\dime S_{\alpha}\}$ and $i_\alpha,j_\alpha,\dotsc \in \{1,\dotsc,\dime F_{\alpha}\}$, and for notational simplicity their collections over the different gauge-group factors are denoted by $a,b,\dotsc$ and $i,j,\dotsc$, respectively.
Moreover, the covariant derivatives of the scalar and fermion fields are defined by
\begin{align}
    (D_\mu\phi)_{a} &= \partial_\mu\phi_{a} -i\sum_{\alpha=1}^{N_G} g_\alpha A^{A_\alpha}_\mu \theta^{A_\alpha}_{ab}\phi_{b}\,,\label{eq:cov_der_phi} \\
    (D_\mu\psi)_{i} &= \partial_\mu\psi_{i} -i\sum_{\alpha=1}^{N_G} g_\alpha A^{A_\alpha}_\mu t^{A_\alpha}_{ij}\psi_{j}\label{eq:cov_der_psi}\,,
\end{align}

\noindent where $g_\alpha$ denote the gauge couplings and $\theta^{A_\alpha}$ and $t^{A_\alpha}$ are the Hermitian generators acting on the scalar and fermion fields, respectively. They satisfy: 
\begin{gather}
    [\theta^{A_\alpha},\theta^{B_\alpha}] = i\, f^{A_\alpha B_\alpha C_\alpha}\,\theta^{C_\alpha} \,, \quad
    [t^{A_\alpha},t^{B_\alpha}] = i\, f^{A_\alpha B_\alpha C_\alpha}\,t^{C_\alpha} \,,
\end{gather}
with the totally anti-symmetric structure constants $f^{A_\alpha B_\alpha C_\alpha}$. The field strength tensors read\footnote{We use the convention $\epsilon^{0123} =-\epsilon_{0123}= 1$.}
\begin{gather}
    F^{A_\alpha}_{\mu\nu}=\partial_\mu A^{A_\alpha}_\nu - \partial_\nu A^{A_\alpha}_\mu + g_\alpha f^{A_\alpha B_\alpha C_\alpha}A^{B_\alpha}_\mu A^{C_\alpha}_\nu\,,\label{eq:field_strength}\\
    \tildee F^{A_\alpha}_{\mu\nu}= \frac{1}{2}\epsilon_{\mu\nu\rho\sigma}  F^{A_\alpha\,\rho\sigma}\,.
\end{gather}
Finally, due to gauge invariance, the gauge kinetic and topological terms take the following form
\begin{equation}
\xi^{A_\alpha B_\beta} = \delta_{\alpha\beta}\delta^{A_\alpha B_\beta}\,,\quad \tildee \xi^{A_\alpha B_\beta} = \frac{\vartheta_\alpha g_\alpha^2}{32\pi^2} \delta_{\alpha\beta}\delta^{A_\alpha B_\beta} \,,
\end{equation}
with the adjoint indices $A_\alpha$ and $B_\beta$ belonging to non-Abelian gauge factors. In the case of abelian $U(1)$ factors that lead to off-diagonal kinetic and topological terms~\cite{Holdom:1985ag,delAguila:1988jz}, the gauge couplings $g_{\alpha}$ need to be generalized to a matrix $g^{A_\alpha B_\beta}$. In order to allow for this possibility, the simple substitution
\begin{equation}
    g_\alpha \theta^{A_\alpha} \to \sum_{\beta=1}^{N_G} g^{A_\alpha B_\beta} \theta^{B_\beta}
\end{equation} 
can be made in the RGEs below,
and similarly with the fermionic generators.

\subsection{Dimension-5 operators}

The dimension-five Lagrangian can again be split into a bosonic and a fermionic part. It takes the form
\begingroup
\interdisplaylinepenalty=10000
\begin{align}
    \mathcal{L}^{(5)} &= \mathcal{L}^{(5)}_{\text{bos}} + \mathcal{L}^{(5)}_{\text{fer}}\,,\label{eq:L5}\\
    \mathcal{L}^{(5)}_{\text{bos}}&=\sum_{\alpha=1}^{N_G}\sum_{\beta=1}^\alpha \left(\left[C_{\phi F^2}\right]^{A_\alpha B_\beta}_{a}\phi_{a}F^{A_\alpha}_{\mu\nu}F^{B_\beta \, \mu\nu}+\left[C_{\phi \tildee F^2}\right]^{A_\alpha B_\beta}_{a}\phi_{a}F^{A_\alpha}_{\mu\nu}\tildee F^{B_\beta \, \mu\nu}\right)
    \nonumber\\&\quad 
    + \left[C_{\phi^5}\right]_{abcde}\phi_{a}\phi_{b}\phi_{c}\phi_{d}\phi_{e}\,,\\
    \mathcal{L}^{(5)}_{\text{fer}} &= \left[C_{\psi^2\phi^2}\right]_{ijab} \psi_i \psi_j \phi_a \phi_b + \sum_{\alpha=1}^{N_G} \left[C_{\psi^2F}\right]_{ij}^{A_\alpha} \psi_i \sigma^{\mu\nu} \psi_j F_{\mu\nu}^{A_\alpha}  + \text{h.c.}\,,
\end{align}
\endgroup\ignorespacesafterend
where $\sigma^{\mu\nu} = \frac{i}{4}(\sigma^\mu \overline \sigma^\nu - \sigma^\nu \overline \sigma^\mu)$ and $\sigma^\mu = (\mathbb{1}_{2\times 2}, \vec \sigma)$. In particular, the fermionic operators exhibit non-trivial symmetry properties due to various Dirac relations, which are equivalently reflected in the corresponding kinematic structures. For instance, for anti-commuting spinor fields one finds:
\begin{gather}
    \psi_i \psi_j = \psi_j\psi_i\,, \quad
    \psi_i \sigma^{\mu\nu} \psi_j = -\psi_j \sigma^{\mu\nu}\psi_i\,,
\end{gather}
and analogously for the right-handed spinors. The symmetries of the dimension-five operators together with their minimal form factors are given in Tab.~\ref{tab:dim5}. 
\begin{table}[tbp]
\renewcommand{\arraystretch}{2}
\begin{align*}
\resizebox{\textwidth}{!}{
\begin{array}[t]{|c|c|c|c|}
\toprule
\text{Name} & \text{Operator} &  \text{Symmetry} & \text{Form factor} \\
\midrule\midrule
\mathcal{O}_{\phi^5} & \phi_a\phi_b\phi_c\phi_d\phi_e & [C_{\phi^5}]_{abcde}=[C_{\phi^5}]_{(abcde)} & F_{\phi^5}(1_{a},2_{b},3_{c},4_{d},5_{e}) = 5! \left[C_{\phi^5}\right]_{abcde}\\
\mathcal{O}_{\phi F^2} & \phi_{a}F^{A_\alpha}_{\mu\nu}F^{B_\beta \, \mu\nu} & \left[C_{\phi F^2}\right]^{A_\alpha B_\beta}_{a}=\left[C_{\phi F^2}\right]^{(A_\alpha B_\beta)}_{a} &
F_{\phi F^2}(1_{a},2^-_{A_\alpha},3^-_{B_\beta}) = - \mathcal S_{\alpha\beta} \left[C_{\phi F^2}\right]^{A_\alpha B_\beta}_{a}\agl{2}{3}^2\\
\mathcal{O}_{\phi \tildee F^2} & \phi_{a}F^{A_\alpha}_{\mu\nu}\tildee F^{B_\beta \, \mu\nu} & \left[C_{\phi \tildee F^2}\right]^{A_\alpha B_\beta}_{a}=\left[C_{\phi \tildee F^2}\right]^{(A_\alpha B_\beta)}_{a} 
 & F_{\phi \tildee F^2}(1_{a},2^-_{A_\alpha},3^-_{B_\beta})  = - i\mathcal S_{\alpha\beta} \left[C_{\phi \tildee F^2}\right]^{A_\alpha B_\beta}_{a}\agl{2}{3}^2\\
 \mathcal{O}_{\psi^2\phi^2} & \psi_i \psi_j \phi_a \phi_b  & \left[C_{\psi^2\phi^2}\right]_{ijab} = \left[C_{\psi^2\phi^2}\right]_{(ij)ab}  = \left[C_{\psi^2\phi^2}\right]_{ij(ab)} 
 & F_{\psi^2\phi^2}(1^-_{i},2^-_{j},3_a,4_b)  = 4 \left[C_{\psi^2\phi^2}\right]_{ijab}\agl{1}{2}\\
 \mathcal{O}_{\psi^2 F} &  \psi_i \sigma^{\mu\nu} \psi_j F_{\mu\nu}^{A_\alpha} & \left[C_{\psi^2 F}\right]_{ij}^{A_\alpha} = \left[C_{\psi^2 F}\right]_{[ij]}^{A_\alpha} 
 & F_{\psi^2 F}(1^-_{i},2^-_{j},3^-_{A_\alpha})  = 2\sqrt{2}\left[C_{\psi^2F}\right]_{ij}^{A_\alpha}\agl{1}{3}\agl{2}{3}\\
\bottomrule
\end{array}
}
\end{align*}
\caption{Dimension-five bosonic and fermionic operators of the general EFT, together with their form factors. The symmetry factors $\mathcal{S}_{\alpha\beta}$ are defined in App.~\ref{sec:symm_factors}.}
\label{tab:dim5}
\end{table}

\subsection{Dimension-6 operators}\label{subsec:dim6_general}
The dimension-six interactions are given by the following Lagrangian
\begingroup
\interdisplaylinepenalty=10000
\begin{align}\label{eq:L6}
    \mathcal{L}^{(6)} &= \mathcal{L}^{(6)}_{\text{bos}}+\mathcal{L}^{(6)}_{\text{fer}}\,,\\
    \mathcal{L}^{(6)}_{\text{bos}}&=\sum_{\alpha=1}^{N_G}\sum_{\beta=1}^\alpha \left(\left[C_{\phi^2 F^2}\right]^{A_\alpha B_\beta}_{ab}\phi_{a}\phi_{b}F^{A_\alpha}_{\mu\nu}F^{B_\beta \, \mu\nu}+\left[C_{\phi^2 \tildee F^2}\right]^{A_\alpha B_\beta}_{ab}\phi_{a}\phi_{b}F^{A_\alpha}_{\mu\nu}\tildee F^{B_\beta \, \mu\nu}\right)
    \nonumber\\&\quad  \nonumber
    + \left[C_{\phi^6}\right]_{abcdef}\phi_{a}\phi_{b}\phi_{c}\phi_{d}\phi_{e}\phi_{f}+\left[C_{D^2\phi^4}\right]_{abcd}(D_\mu \phi)_{a}(D^\mu \phi)_{b}\phi_{c}\phi_{d}\\&\quad
    +\sum_{\alpha=1}^{N_G}\left(\left[C_{F^3}\right]^{A_\alpha B_\alpha C_\alpha}F^{A_\alpha\,\nu}_{\mu}F^{B_\alpha\,\rho}_{ \nu} F^{C_\alpha\,\mu}_{\rho}+\left[C_{\widetilde F^3}\right]^{A_\alpha B_\alpha C_\alpha}F^{A_\alpha\,\nu}_{\mu}F^{B_\alpha\,\rho}_{ \nu} \widetilde F^{C_\alpha\,\mu}_{\rho}\right)\,,\\
    \mathcal{L}^{(6)}_{\text{fer}} &= \bigg(\left[C_{\psi^2\phi^3}\right]_{ijabc} \psi_i \psi_j \phi_a \phi_b \phi_c + \sum_{\alpha=1}^{N_G} \left[C_{\psi^2 \phi F}\right]_{ija}^{A_\alpha} \psi_i \sigma^{\mu\nu} \psi_j \phi_a F_{\mu\nu}^{A_\alpha}  \nonumber \\&\quad + \left[C_{\psi^4}\right]_{ijkl} (\psi_i \psi_j)(\psi_k\psi_l) + \text{h.c.}\bigg) + \left[C_{\overline \psi^2 \psi^2}\right]_{ijkl}(\psi_i^\dagger \overline \sigma^\mu \psi_j)(\psi_k^\dagger \overline \sigma_\mu \psi_l)\nonumber \\&\quad
    + i\left[C_{D\overline \psi \psi \phi^2}\right]_{ijab}(\psi^\dagger_i \overline \sigma^\mu \psi_j)\left[(D_\mu \phi)_a \phi_b - \phi_a (D_\mu \phi)_b\right]\,.
\end{align}
\endgroup\ignorespacesafterend
The symmetries of the corresponding Wilson coefficients are collected for bosonic and fermionic operators in Tab.~\ref{tab:dim6}, and the minimal form factors are given in Tab.~\ref{tab:dim6FFs}. For instance, the four-fermion operator $\mathcal{O}_{\overline \psi^2 \psi^2}$ satisfies the following Fierz relation:
\begin{equation}
    (\psi_i^\dagger \overline \sigma_\mu \psi_j)(\psi_k^\dagger \overline \sigma^\mu \psi_l) = 2 (\psi_i^\dagger \psi_k^\dagger)( \psi_l \psi_j) \,,
\end{equation}
which leads to several symmetry relations. Furthermore, using the identity
\begin{equation}
    (\psi_i^\dagger \overline \sigma_\mu \psi_j) = - (\psi_j \sigma_\mu \psi_i^\dagger)\,,
\end{equation}
the operator $\mathcal{O}_{\overline \psi^2 \psi^2}$ describes four-fermion vector operators with both equal and mixed chiralities. 
Moreover, as a consequence of the Schouten identity 
\begin{equation}
    \agl{1}{2}\agl{3}{4} + \agl{1}{3}\agl{4}{2}+\agl{1}{4}\agl{2}{3} = 0 \,,
\end{equation}
 the Wilson coefficient associated with the operator $\mathcal O_{\psi^4}$ satisfies
\begin{equation}
    \left[C_{\psi^4}\right]_{ijkl} + \left[C_{\psi^4}\right]_{iklj} + \left[C_{\psi^4}\right]_{iljk} = 0\,,
\end{equation}
in addition to the identities reported in Tab.~\ref{tab:dim6}.

\begin{table}[tbp]
\renewcommand{\arraystretch}{2.3}
\begin{align*}
\resizebox{\textwidth}{!}{
\begin{array}[t]{|c|c|c|c|}
\toprule
\text{Name} & \text{Operator} &  \text{Symmetry} \\
\midrule\midrule
\mathcal O_{\phi^6} & \phi_a\phi_b\phi_c\phi_d\phi_e\phi_f & [C_{\phi^6}]_{abcdef}=[C_{\phi^6}]_{(abcdef)} \\
\mathcal O_{D^2\phi^4}  & (D_\mu \phi)_{a}(D^\mu \phi)_{b}\phi_{c}\phi_{d} & [C_{D^2\phi^2}]_{abcd} = [C_{D^2\phi^2}]_{(ab)cd} = [C_{D^2\phi^2}]_{ab(cd)}
\\
\mathcal{O}_{\phi^2 F^2} & \phi_{a}\phi_{b}F^{A_\alpha}_{\mu\nu}F^{B_\beta \, \mu\nu} & \left[C_{\phi^2F^2}\right]^{A_\alpha B_\beta}_{ab}=\left[C_{\phi^2F^2}\right]^{(A_\alpha B_\beta)}_{ab}=\left[C_{\phi^2F^2}\right]^{A_\alpha B_\beta}_{(ab)} \\
\mathcal{O}_{\phi^2\tildee F^2} & \phi_{a}\phi_{b}F^{A_\alpha}_{\mu\nu}\tildee F^{B_\beta \, \mu\nu} & \left[C_{\phi^2 \tildee F^2}\right]^{A_\alpha B_\beta}_{ab}=\left[C_{\phi^2\tildee F^2}\right]^{(A_\alpha B_\beta)}_{ab}=\left[C_{\phi^2\tildee F^2}\right]^{A_\alpha B_\beta}_{(ab)}\\
 \mathcal{O}_{F^3} & F^{A_\alpha\,\nu}_{\mu}F^{B_\alpha\,\rho}_{ \nu}F^{C_\alpha\,\mu}_{\rho} & \left[C_{F^3}\right]^{A_\alpha B_\alpha C_\alpha}=\left[C_{F^3}\right]^{[A_\alpha B_\alpha C_\alpha]}\\
  \mathcal{O}_{\tildee F^3} & F^{A_\alpha\,\nu}_{\mu}F^{B_\alpha\,\rho}_{ \nu}{\tildee F}^{C_\alpha\,\mu}_{\rho} & \left[C_{\tildee F^3}\right]^{A_\alpha B_\alpha C_\alpha}=\left[C_{\tildee F^3}\right]^{[A_\alpha B_\alpha C_\alpha]}\\
\midrule
\mathcal O_{\psi^2\phi^3} &  \psi_i \psi_j \phi_a \phi_b \phi_c & \left[C_{\psi^2\phi^3}\right]_{ijabc} = \left[C_{\psi^2\phi^3}\right]_{ij(abc)} = \left[C_{\psi^2\phi^3}\right]_{(ij)abc} \\
\mathcal O_{\psi^2 \phi F}  & \psi_i \sigma^{\mu\nu} \psi_j \phi_a F_{\mu\nu}^{A_\alpha} & \left[C_{\psi^2 \phi F}\right]_{ija}^{A_\alpha} = \left[C_{\psi^2 \phi F}\right]_{[ij]a}^{A_\alpha}
\\
\mathcal{O}_{\psi^4} & (\psi_i \psi_j)(\psi_k\psi_l) & \left[C_{\psi^4}\right]_{ijkl} = \left[C_{\psi^4}\right]_{(ij)kl} = \left[C_{\psi^4}\right]_{ij(kl)} = \left[C_{\psi^4}\right]_{klij}  \\
\mathcal{O}_{\overline \psi^2 \psi^2} & (\psi_i^\dagger \overline \sigma^\mu \psi_j)(\psi_k^\dagger \overline \sigma_\mu \psi_l) & \left[C_{\overline \psi^2 \psi^2}\right]_{ijkl} = \left[C_{\overline \psi^2 \psi^2}\right]_{kjil} = \left[C_{\overline \psi^2 \psi^2}\right]_{ilkj} = \left[C_{\overline \psi^2 \psi^2}\right]_{jilk}^*\\
 \mathcal{O}_{D\overline \psi \psi \phi^2} & i(\psi^\dagger_i \overline \sigma^\mu \psi_j)\left[(D_\mu \phi)_a \phi_b - \phi_a (D_\mu \phi)_b\right] & \left[C_{D\overline \psi \psi \phi^2}\right]_{ijab} = \left[C_{D\overline \psi \psi \phi^2}\right]_{ij[ab]} = \left[C_{D\overline \psi \psi \phi^2}\right]_{jiba}^*\\
\bottomrule
\end{array}
}
\end{align*}
\caption{Dimension-six bosonic and fermionic operators of the general EFT together with the corresponding Wilson coefficients.}
\label{tab:dim6}
\end{table}

\begin{table}[tbp]
\renewcommand{\arraystretch}{2.5}
\begin{align*}
\resizebox{\textwidth}{!}{
\begin{array}[t]{|c|c|c|c|}
\toprule
\text{Name}  & \text{Form factor} \\
\midrule\midrule
\mathcal O_{\phi^6} & F_{\phi^5}(1_{a},2_{b},3_{c},4_{d},5_{e},6_{f}) = 6! \left[C_{\phi^6}\right]_{abcdef}\\
\mathcal O_{D^2\phi^4}  & F_{D^2\phi^4}(1_a,2_b,3_c,4_d) =-2\left(\left[\widehat C_{D^2\phi^4}\right]_{abcd} s_{12}  + \left[\widehat C_{D^2\phi^4}\right]_{acbd} s_{13}\right)\\
\mathcal{O}_{\phi^2 F^2} &
F_{\phi^2 F^2}(1_a,2_b,3^-_{A_\alpha},4^-_{B_\beta}) = -2 \mathcal S_{\alpha\beta} \left[C_{\phi^2 F^2}\right]^{A_\alpha B_\beta}_{a b}\agl{3}{4}^2\\
\mathcal{O}_{\phi^2\tildee F^2} & F_{\phi^2 \tildee F^2}(1_a,2_b,3^-_{A_\alpha},4^-_{B_\beta}) = -2i \mathcal S_{\alpha\beta} \left[C_{\phi^2 \tildee F^2}\right]^{A_\alpha B_\beta}_{a b}\agl{3}{4}^2\\
 \mathcal{O}_{F^3} & F_{F^3}(1^-_{A_\alpha},2^-_{B_\alpha},3^-_{C_\alpha}) =-3i\sqrt{2} \left[C_{F^3}\right]^{A_\alpha B_\alpha C_\alpha} \agl{1}{2}\agl{2}{3}\agl{3}{1}\\
  \mathcal{O}_{\tildee F^3} & F_{\tildee F^3}(1^-_{A_\alpha},2^-_{B_\alpha},3^-_{C_\alpha}) =3\sqrt{2} \left[C_{\tildee F^3}\right]^{A_\alpha B_\alpha C_\alpha} \agl{1}{2}\agl{2}{3}\agl{3}{1}\\
 \midrule
\mathcal O_{\psi^2\phi^3} & F_{\psi^2 \phi^3}(1^-_i,2^-_j,3_a,4_b,5_c) = 12 \left[ C_{\psi^2\phi^3}\right]_{ijabc} \agl{1}{2}\\
\mathcal O_{\psi^2 \phi F}   & F_{\psi^2 \phi F}(1^-_i,2^-_j,3_a,4^-_{A_\alpha}) = 2\sqrt{2} \left[C_{\psi^2 \phi F}\right]_{ija}^{A_\alpha}\agl{1}{4}\agl{2}{4}\\
\mathcal{O}_{\psi^4} &
F_{\psi^4}(1^-_i,2^-_j,3^-_k,4^-_l) = 8 \left(\left[C_{\psi^4}\right]_{ijkl}-\left[C_{\psi^4}\right]_{iljk}\right)\agl{1}{2}\agl{3}{4} + 8 \left(\left[C_{\psi^4}\right]_{ikjl}-\left[C_{\psi^4}\right]_{iljk}\right)\agl{1}{3}\agl{4}{2} \\
\mathcal{O}_{\overline \psi^2 \psi^2} & F_{\overline \psi^2 \psi^2}(1^+_i,2^-_j,3^+_k,4^-_l) = 8 \left[C_{\overline \psi^2 \psi^2}\right]_{ijkl}\agl{2}{4}\sqr{3}{1}\\
\mathcal{O}_{D\overline \psi \psi \phi^2} & F_{D\overline \psi \psi \phi^2}(1^+_i,2^-_j,3_a,4_b) = -4 \left[C_{D\overline \psi \psi \phi^2}\right]_{ijab}\agl{2}{3}\sqr{3}{1}\\
\bottomrule
\end{array}
}
\end{align*}
\caption{Dimension-six bosonic and fermionic operators and their form factors. $s_{ij} = \agl{i}{j}\sqr{j}{i}$ are the Mandelstam invariants. The symmetry factors $\mathcal{S}_{\alpha\beta}$ are defined in App.~\ref{sec:symm_factors}.}
\label{tab:dim6FFs}
\end{table}

Let us further note a subtlety that involves the $D^2\phi^4$-class operator $[\mathcal O_{D^2\phi^4}]_{abcd} = (D_\mu \phi)_a (D^\mu \phi)_b \phi_c \phi_d$. 
For this, we define the following combination of Wilson coefficients:
\begin{equation}\label{eq:hatCD2phi4}
\left[\widehat C_{D^2\phi^4}\right]_{abcd}=\left[C_{D^2\phi^4}\right]_{abcd}+\left[C_{D^2\phi^4}\right]_{cdab}-\left[C_{D^2\phi^4}\right]_{adbc}-\left[C_{D^2\phi^4}\right]_{bcad}\,.
\end{equation}
Adopting the geometric approach this combination can be shown to be proportional to the Riemann tensor of the scalar metric,\footnote{For a scalar metric $g_{ab}(\phi) = \delta_{ab} + 2[C_{D^2\phi^4}]_{abcd}\phi^c \phi^d$, the corresponding Riemann tensor $R_{abcd}$ is proportional to $[\widehat C_{D^2\phi^4}]_{acbd} = [C_{D^2\phi^4}]_{acbd}+[C_{D^2\phi^4}]_{bdac}-[C_{D^2\phi^4}]_{adbc}-[C_{D^2\phi^4}]_{bcad}$.} which implies the following symmetries
\begin{gather}
    \left[\widehat C_{D^2\phi^4}\right]_{abcd} = \left[\widehat C_{D^2\phi^4}\right]_{badc} = -\left[\widehat C_{D^2\phi^4}\right]_{adcb} = -\left[\widehat C_{D^2\phi^4}\right]_{cbad}\,,\\
    \left[\widehat C_{D^2\phi^4}\right]_{abcd} + \left[\widehat C_{D^2\phi^4}\right]_{acdb} + \left[\widehat C_{D^2\phi^4}\right]_{adbc} = 0\,.
\end{gather}
Furthermore, introducing the redundant operator

\begin{equation}
\left[\mathcal R_{D^2\phi^4}\right]_{abcd} = (\Box \phi)_a \phi_b \phi_c \phi_d\,,
\end{equation}
and using integration by parts, one finds the relations
\begin{align}
    \left[\mathcal O_{D^2\phi^4}\right]_{abcd} - \left[\mathcal O_{D^2\phi^4}\right]_{cdab} &= -\frac{1}{2}\left(
    \left[\mathcal R_{D^2\phi^4}\right]_{abcd} + \left[\mathcal R_{D^2\phi^4}\right]_{bacd}\right.\nonumber\\
    &\left.\qquad\,\,-\left[\mathcal R_{D^2\phi^4}\right]_{cabd}-\left[\mathcal R_{D^2\phi^4}\right]_{dabc}\,
    \right)
\end{align}
and
\begin{equation}
    \left[\mathcal O_{D^2\phi^4}\right]_{abcd} + \left[\mathcal O_{D^2\phi^4}\right]_{adbc} +
    \left[\mathcal O_{D^2\phi^4}\right]_{acdb}  = -\left[\mathcal R_{D^2\phi^4}\right]_{abcd}\,,\label{eq:D2phi4symmetric}
\end{equation}
implying
\begin{equation}
\left[\mathcal O_{D^2\phi^4}\right]_{(abcd)} = -\frac{1}{3}\left[\mathcal R_{D^2\phi^4}\right]_{(abcd)}\,.
\end{equation}
One can then show, using the equation of motion of the scalar field
\begin{equation}
    (\Box \phi)_a = -t_a -m^2_{ab}\,\phi_b - \frac{1}{2!}h_{abc}\,\phi_b\phi_c - \frac{1}{3!}\lambda_{abcd}\,\phi_b\phi_c\phi_d
    -\frac{1}{2!}\left(y_{ija}\psi_i \psi_j + y_{ija}^* \psi_i^\dagger \psi_j^\dagger
    \right)
    \,,
\end{equation}
that the redundant operator $\mathcal R_{D^2\phi^4}$ is a linear combination of $\mathcal O_{\phi^6}$, $\mathcal O_{\phi^5}$, $\mathcal O_{\phi^4}$, $\mathcal O_{\phi^3}$, and $\mathcal O_{\psi^2\phi^3}$.
Consequently, generating $\mathcal R_{D^2\phi^4}$ at the one-loop level influences the RGEs of the parameters $C_{\phi^6}$, $C_{\phi^5}$, $\lambda$, $h$, and $C_{\psi^2\phi^3}$.
To illustrate this point we write the Wilson coefficient $[C_{D^2\phi^4}]_{abcd}$ as
\begin{equation}
    \left[C_{D^2\phi^4}\right]_{abcd} = \frac{1}{6}\left(
    2\left[\widehat C_{D^2\phi^4}\right]_{abcd}-\left[\widehat C_{D^2\phi^4}\right]_{acbd}
    \right)+\left[\tildee C_{D^2\phi^4}\right]_{abcd}+\left[\overline C_{D^2\phi^4}\right]_{abcd}\,,
    \label{eq:decomposition}
\end{equation}
where we introduced the combinations
\begin{align}
    \left[\tildee C_{D^2\phi^4}\right]_{abcd} &= \left[ C_{D^2\phi^4}\right]_{(abcd)}\,,\label{eq:Ctilde_def}\\
    \left[\overline C_{D^2\phi^4}\right]_{abcd} &= \frac{1}{2}\left(
    \left[C_{D^2\phi^4}\right]_{abcd} - \left[C_{D^2\phi^4}\right]_{cdab}
    \right)\,.\label{eq:Cbar_def}
\end{align}
In this notation, one finds the following contributions from $\mathcal R_{D^2\phi^4}$
\begin{align}
    \left[\dot C_{\psi^2\phi^3}\right]_{ijabc} &\supset \frac{1}{3!}\frac{1}{2!}\sum_{\sigma(\{a,b,c\})}y_{ijd} \left(
    \frac{1}{3}\left[\dot{\tildee C}_{D^2\phi^4}\right]_{dabc}+\left[\dot{\overline C}_{D^2\phi^4}\right]_{dabc}
    \right)\,,\label{eq:psi2phi3<-Ctildeandbar}\\
    \left[\dot C_{\phi^6}\right]_{abcdef} &\supset \frac{1}{6!}\frac{1}{3!}\sum_{\sigma(\{a,b,c,d,e,f\})}\lambda_{abcg}\left(
    \frac{1}{3}\left[\dot{\tildee C}_{D^2\phi^4}\right]_{gdef}+\left[\dot{\overline C}_{D^2\phi^4}\right]_{gdef}
    \right)\,,\label{eq:phi6<-Ctildeandbar}\\
    \left[\dot C_{\phi^5}\right]_{abcde} &\supset \frac{1}{5!}\frac{1}{2!}\sum_{\sigma(\{a,b,c,d,e\})}h_{abf}\left(
    \frac{1}{3}\left[\dot{\tildee C}_{D^2\phi^4}\right]_{fcde}+\left[\dot{\overline C}_{D^2\phi^4}\right]_{fcde}
    \right)\,,\label{eq:phi5<-Ctildeandbar}\\
    \dot \lambda_{abcd} &\supset - \sum_{\sigma(\{a,b,c,d\})}m^2_{ae}\left(
    \frac{1}{3}\left[\dot{\tildee C}_{D^2\phi^4}\right]_{ebcd}+\left[\dot{\overline C}_{D^2\phi^4}\right]_{ebcd}
    \right)\,,\label{eq:phi4<-Ctildeandbar}\\
    \dot h_{abc} &\supset -  \sum_{\sigma(\{a,b,c\})} t_d \left(
    \frac{1}{3}\left[\dot{\tildee C}_{D^2\phi^4}\right]_{dabc}+\left[\dot{\overline C}_{D^2\phi^4}\right]_{dabc}
    \right)\,,\label{eq:phi3<-Ctildeandbar}
\end{align}
which have to be taken into account and added to the results presented in Sec.~\ref{sec:results}.
With $\sum_{\sigma(I)}$ we denote the sum over all the permutations of the set of indices $I$, and $\sum_{\sigma(I\times J)} = \sum_{\sigma(I)}\sum_{\sigma(J)}$.

\section{Review of the method}
\label{sec:method}

In this section we briefly review the method used to compute the complete set of one-loop RGEs. For more details and explicit examples of calculations we refer to \cite{Aebischer:2025zxg,Bresciani:2024shu}. 
A key ingredient for the computation is the form factor $F$, which is defined as a matrix element of the operator $\mathcal O_i$ between the vacuum and an on-shell $n$-particle state $\ket{\vec n}$:
\begin{equation}
    F_i(\vec{n};q) =  \mel{\vec n}{\mathcal O_i(q)}{0}\,,
\end{equation}
where $q$ denotes the off-shell momentum injected by the operator.
Using fundamental principles such as analyticity, unitarity, and the CPT theorem, together with the Callan-Symanzik equation, one can derive at one-loop order \cite{Caron-Huot:2016cwu,Bresciani:2023jsu,EliasMiro:2021jgu}
\begin{equation}\label{eq:FFRGE}
\left(\gamma_{i\leftarrow j}-\delta_{ij}\gamma_{i,\text{IR}}\right) F_i|_* = -\frac{1}{\pi} (\mathcal M F_j)|_*\,,
    \qquad 
    \gamma_{i\leftarrow j,k} F_i|_* = -\frac{1}{\pi}\left.\frac{\partial }{\partial C_k} \right|_* (\mathcal M F_j)\,,
\end{equation}
which relate form factors to the anomalous dimensions governing the renormalization of the Wilson coefficients $C_i$ for single and double operator insertions. 
Here, the symbol ``$*$'' denotes the Gaussian fixed point, at which all Wilson coefficients vanish. 
Ultraviolet anomalous dimensions and $\beta$-functions are linked through
\begin{equation}
    \beta_i(\{C_k\}) = \mu \dv[]{C_i}{\mu} =\frac{1}{16\pi^2} \dot C_i = \sum_{n>0}\frac{1}{n!}\gamma_{i\leftarrow j_1,\dotsc,j_n} C_{j_1}\dotsc C_{j_n}\,,
\end{equation}
while $\gamma_{i,\text{IR}}$ denotes the infrared anomalous dimension associated with the external states of the form factor, arising from soft and collinear emissions~\cite{Sterman:2002qn,Becher:2009cu,Chiu:2009mg}.
The collinear contributions are collected in App.~\ref{sec:coll_ADM}, whereas the logarithmic soft contributions are automatically identified and removed by employing the integration method based on Stokes theorem \cite{Mastrolia:2009dr}.
The integrals under consideration correspond to the two-particle unitarity cuts between tree-level amplitudes and form factors that are present on the right-hand sides of Eq.~\eqref{eq:FFRGE}:
\begin{equation}
    (\mathcal M F_j)(1,\dots,n) = 
   \sum_{k=2}^n \sum_{\{x,y\}}\int \text{dLIPS}_2
   \sum_{h_1,h_2} F_j(x^{h_1},y^{h_2},k + 1,\dots,n) \mathcal M(1,\dots,k;x^{h_1},y^{h_2})\,,
\end{equation}
where $\mathcal M(\vec n;\vec m) = \mel{\vec n}{\mathcal M}{\vec m}$.
We used the package \texttt{S@M} \cite{Maitre:2007jq} to handle spinor-helicity expressions and perform the integrals.

\section{Results}
\label{sec:results}
This section contains the complete one-loop RGEs for the fermionic operator sector of the general EFT, including the fermion-induced contributions to the running of bosonic operators. The complete set of purely bosonic RGEs can be found in \cite{Aebischer:2025zxg}.

\subsection{Running of dimension-6 operators}

\subsubsection{\texorpdfstring{$\phi^6$}{phi6} class}

In addition to the results presented below, one needs to supplement the RGEs of the $\phi^6$-class Wilson coefficients with the expressions given in Eq.~\eqref{eq:phi6<-Ctildeandbar}.
The explicit expressions for $\dot{\tildee C}_{D^2\phi^4}$ and $\dot{\overline C}_{D^2\phi^4}$ can be obtained using their definitions in Eqs.~\eqref{eq:Ctilde_def} and \eqref{eq:Cbar_def} and the RGEs for $C_{D^2\phi^4}$ reported in Sec.~\ref{sec:D2phi4RGE}.

\paragraph{$\phi^6 \leftarrow \phi^6$:}

\begin{equation}
    \left[\dot C_{\phi^6}\right]_{abcdef} = \frac{1}{5!} \sum_{\sigma(\{a,b,c,d,e,f\})} \left.\gamma_{c,s}^{fg} \right|_{\text{fer}}\left[C_{\phi^6}\right]_{abcdeg} \,,
\end{equation}
where $\left.\gamma_{c,s}^{ab} \right|_{\text{fer}} = \frac{1}{2}(y_{ija} y_{ijb}^* + y_{ijb} y_{ija}^*)$ is the fermionic part of the collinear anomalous dimension of scalar fields in Eq.~\eqref{eq:coll_scalar}.

\paragraph{\texorpdfstring{$\phi^6\leftarrow \psi^2\phi^3$}{phi6frompsi2phi3}:}

\begin{align}
    \left[\dot{C}_{\phi^6}\right]_{abcdef} &= - \frac{8}{6!}\sum_{\sigma(\{a,b,c,d,e,f\})} \Re{y_{ija}\, y_{jkb}^*\, y_{klc}\left[C_{\psi^2\phi^3}\right]_{lidef}^*}\,.\label{phi6_psi2phi3}
\end{align}

\paragraph{\texorpdfstring{$\phi^6\leftarrow \psi^2\phi^2\times\psi^2\phi^2$}{phi6frompsi2phi2xpsi2phi2}:}

\begin{align}
    \left[\dot C_{\phi^6}\right]_{abcdef} &= \frac{8}{6!} \sum_{\sigma(\{a,b,c,d,e,f\})}\Bigg(\Re{ y^*_{jkc} y^*_{lif} \left[C_{\psi^2\phi^2}\right]_{ijab} \left[C_{\psi^2\phi^2}\right]_{klde}}\\ \nonumber
     &+2 \left[C_{\psi^2\phi^2}\right]_{ijab} y^*_{jkc} \, y_{kld} \left[C_{\psi^2\phi^2}\right]^*_{lief}\Bigg)\,.
     \label{phi6_psi2phi2xpsi2phi2}
\end{align}

\paragraph{\texorpdfstring{$\phi^6\leftarrow \psi^2\phi^2\times\phi^5$}{phi6frompsi2phi2xphi5}:}
\begin{equation}
    \left[\dot{C}_{\phi^6}\right]_{abcdef} = -\frac{1}{36} \sum_{\sigma(\{a,b,c,d,e,f\})} \Re{
      y_{ijg} \left[C_{\phi^5}\right]_{abcdg} \left[C_{\psi^2\phi^2}\right]^*_{jief}}
     \label{phi6_psi2phi2xphi5}\,.
\end{equation}

\subsubsection{\texorpdfstring{$D^2 \phi^4$}{D2phi4} class\label{sec:D2phi4RGE}}

\paragraph{$D^2\phi^4 \leftarrow D^2\phi^4$:}
~
\newline The running of $\widehat C_{D^2\phi^4}$, defined in Eq.~\eqref{eq:hatCD2phi4}, is given by 
\begin{align}
    \left[\dot{\widehat C}_{D^2\phi^4}\right]_{abcd} &=\left.\gamma_{c,s}^{ae}\right|_{\text{fer}}\left[\widehat C_{D^2\phi^4}\right]_{ebcd}+\left.\gamma_{c,s}^{be}\right|_{\text{fer}}\left[\widehat C_{D^2\phi^4}\right]_{aecd}+\left.\gamma_{c,s}^{ce}\right|_{\text{fer}}\left[\widehat C_{D^2\phi^4}\right]_{abed}\nonumber \\
    &+\left.\gamma_{c,s}^{de}\right|_{\text{fer}}\left[\widehat C_{D^2\phi^4}\right]_{abce}\,,
\end{align}
where $\left.\gamma_{c,s}^{ab} \right|_{\text{fer}} = \frac{1}{2}(y_{ija} y_{ijb}^* + y_{ijb} y_{ija}^*)$ is the fermionic part of the collinear anomalous dimension of scalar fields in Eq.~\eqref{eq:coll_scalar}.
Similarly, the running of the full operator (including off-shell contributions) is given by
\begin{align}
    \left[\dot{ C}_{D^2\phi^4}\right]_{abcd} &=\left.\gamma_{c,s}^{ae}\right|_{\text{fer}}\left[ C_{D^2\phi^4}\right]_{ebcd}+\left.\gamma_{c,s}^{be}\right|_{\text{fer}}\left[ C_{D^2\phi^4}\right]_{aecd}+\left.\gamma_{c,s}^{ce}\right|_{\text{fer}}\left[ C_{D^2\phi^4}\right]_{abed}\nonumber \\
    &+\left.\gamma_{c,s}^{de}\right|_{\text{fer}}\left[ C_{D^2\phi^4}\right]_{abce}\,.
\end{align}

\paragraph{\texorpdfstring{$D^2 \phi^4 \leftarrow \psi^2 \phi^2 \times \psi^2 \phi^2 $}{D2phi4frompsi2phi2xpsi2phi2}:}
~
\newline The running of $\widehat C_{D^2\phi^4}$, defined in Eq.~\eqref{eq:hatCD2phi4}, is given by
\begin{align}
    \left[\dot{\widehat C}_{D^2\phi^4}\right]_{abcd} &= 
    8\left[C_{\psi^2\phi^2}\right]_{ijab} \left[C_{\psi^2\phi^2}\right]^*_{jicd} - (b \leftrightarrow d) - (a \leftrightarrow c) + \binom{a \leftrightarrow c}{b \leftrightarrow d} \,,
\end{align}
while the running of the full operator (including off-shell contributions) is given by
\begin{equation}
    \left[\dot C_{D^2\phi^4}\right]_{abcd} = 8\left[C_{\psi^2\phi^2}\right]_{ijab} \left[C_{\psi^2\phi^2}\right]^*_{jicd}\,.
\end{equation}

\paragraph{$D^2 \phi^4 \leftarrow D \overline \psi \psi \phi^2$:}
~
\newline The running of $\widehat C_{D^2\phi^4}$, defined in Eq.~\eqref{eq:hatCD2phi4}, is given by
\begin{align}
    \left[\dot{\widehat C}_{D^2\phi^4}\right]_{abcd} &= 
    \sum_{\sigma(\{a,b\}\times \{c,d\})}\left(
    y_{ikd}y^*_{kjb} - y_{ikb}y^*_{kjd}+\frac{2}{3}\sum_{\alpha=1}^{N_G}g_\alpha^2 t^{A_\alpha}_{ji}\theta^{A_\alpha}_{bd}
    \right) \left[C_{D\overline \psi \psi \phi^2}\right]_{ijac}\nonumber \\
    &-  (b \leftrightarrow d) - (a \leftrightarrow c) + \binom{a \leftrightarrow c}{b \leftrightarrow d}
    \,,
\end{align}
while the running of the full operator (including off-shell contributions) is given by
\begin{align}
    \left[\dot C_{D^2\phi^4}\right]_{abcd} = 
    \sum_{\sigma(\{a,b\}\times \{c,d\})}\left(
    y_{ikd}y^*_{kjb} - y_{ikb}y^*_{kjd}+\frac{2}{3}\sum_{\alpha=1}^{N_G}g_\alpha^2 t^{A_\alpha}_{ji}\theta^{A_\alpha}_{bd}
    \right) \left[C_{D\overline \psi \psi \phi^2}\right]_{ijac}
    \,.
\end{align}

\paragraph{$D^2\phi^4 \leftarrow \psi^2 F \times \psi^2 F$:}
~
\newline The running of $\widehat C_{D^2\phi^4}$, defined in Eq.~\eqref{eq:hatCD2phi4}, is given by
\begin{align}
    \left[\dot{\widehat C}_{D^2\phi^4}\right]_{abcd} &= -\frac{1}{3} \sum_{\alpha=1}^{N_G} \sum_{\beta=1}^{N_G} g_\alpha g_\beta \left(\theta^{A_\alpha}_{ad}\theta^{B_\beta}_{bc} + \theta^{A_\alpha}_{ac}\theta^{B_\beta}_{bd}\right)\nonumber \\
    &\times 
    \left(
    \left[C_{\psi^2 F}\right]_{ij}^{A_\alpha *} \left[C_{\psi^2 F}\right]_{ji}^{B_\beta} + \left[C_{\psi^2 F}\right]_{ij}^{B_\beta *} \left[C_{\psi^2 F}\right]_{ji}^{A_\alpha}
    \right) \nonumber \\&-  (b \leftrightarrow d) - (a \leftrightarrow c) + \binom{a \leftrightarrow c}{b \leftrightarrow d}\,,
\end{align}
while the running of the full operator (including off-shell contributions) is given by
\begin{align}
    \left[\dot C_{D^2\phi^4}\right]_{abcd} &= -\frac{1}{3} \sum_{\alpha=1}^{N_G} \sum_{\beta=1}^{N_G} g_\alpha g_\beta \left(\theta^{A_\alpha}_{ad}\theta^{B_\beta}_{bc} + \theta^{A_\alpha}_{ac}\theta^{B_\beta}_{bd}\right)\nonumber \\
    &\times 
    \left(
    \left[C_{\psi^2 F}\right]_{ij}^{A_\alpha *} \left[C_{\psi^2 F}\right]_{ji}^{B_\beta} + \left[C_{\psi^2 F}\right]_{ij}^{B_\beta *} \left[C_{\psi^2 F}\right]_{ji}^{A_\alpha}
    \right)\,.
\end{align}

\subsubsection{\texorpdfstring{$\phi^2F^2$}{phi2F2} class}

\paragraph{$\phi^2 F^2 \leftarrow \phi^2 F^2$:}

\begin{align}
    \left[\dot{C}_{\phi^2 F^2}\right]_{ab}^{A_\alpha B_\beta} &= \sum_{\sigma(\{a,b\})}\left.\gamma_{c,s}^{ac}\right|_{\text{fer}}\left[ C_{\phi^2  F^2}\right]_{cb}^{A_\alpha B_\beta} + \left.\gamma^{A_\alpha C_\alpha}_{c,v}\right|_{\text{fer}}\left[ C_{\phi^2  F^2}\right]_{ab}^{C_\alpha B_\beta} \nonumber \\
    &+ \left.\gamma^{B_\beta C_\beta}_{c,v}\right|_{\text{fer}}\left[ C_{\phi^2  F^2}\right]_{ab}^{A_\alpha C_\beta} \,,
    \\
    \left[\dot{C}_{\phi^2 \widetilde F^2}\right]_{ab}^{A_\alpha B_\beta} &= \sum_{\sigma(\{a,b\})}\left.\gamma_{c,s}^{ac}\right|_{\text{fer}}\left[ C_{\phi^2  \widetilde F^2}\right]_{cb}^{A_\alpha B_\beta} + \left.\gamma^{A_\alpha C_\alpha}_{c,v}\right|_{\text{fer}}\left[ C_{\phi^2  \widetilde F^2}\right]_{ab}^{C_\alpha B_\beta} \nonumber \\
    &+ \left.\gamma^{B_\beta C_\beta}_{c,v}\right|_{\text{fer}}\left[ C_{\phi^2  \widetilde F^2}\right]_{ab}^{A_\alpha C_\beta}\,,
\end{align}
where $\left.\gamma_{c,s}^{ab} \right|_{\text{fer}} = \frac{1}{2}(y_{ija} y_{ijb}^* + y_{ijb} y_{ija}^*)$ is the fermionic part of the collinear anomalous dimension of scalar fields in Eq.~\eqref{eq:coll_scalar} and $\left.\gamma^{A_\alpha B_\alpha}_{c,v}\right|_{\text{fer}} = \frac{2}{3}g_\alpha^2 \Tr(t^{A_\alpha }t^{B_\alpha})$ the fermionic part of the gauge bosons collinear anomalous dimension in Eq.~\eqref{eq:coll_vector}.

\paragraph{$\phi^2F^2 \leftarrow \psi^2 \phi F$:}

\begin{align}
    \left[\dot C_{\phi^2 F^2}\right]_{ab}^{A_\alpha B_\beta} &= 2 \mathcal S_{\alpha\beta}^{-1} \sum_{\sigma(\{A_\alpha,B_\beta\} \times \{a,b\})}
     g_\beta \Re{
     y^*_{ikb}t^{B_\beta}_{jk}
    \left[C_{\psi^2\phi F}\right]_{ija}^{A_\alpha}
    }
    \,,\label{phi2F2_psi2phiF}
    \\
    \left[\dot C_{\phi^2 \widetilde F^2}\right]_{ab}^{A_\alpha B_\beta} &= 2 \mathcal S_{\alpha\beta}^{-1} \sum_{\sigma(\{A_\alpha,B_\beta\} \times \{a,b\})}
     g_\beta \Im{
     y^*_{ikb}t^{B_\beta}_{jk}
    \left[C_{\psi^2\phi F}\right]_{ija}^{A_\alpha}
    }
    \,.\label{phi2Ftilde2_psi2phiF}
\end{align}

\paragraph{$\phi^2 F^2\leftarrow \psi^2 F \times \psi^2 F$:}

\begin{align}
    \left[\dot C_{\phi^2 F^2}\right]_{ab}^{A_\alpha B_\beta} &= 8 \mathcal S_{\alpha\beta}^{-1} \Re{y^*_{jia} \left[C_{\psi^2 F}\right]_{il}^{A_\alpha} y^*_{lkb} \left[C_{\psi^2 F}\right]_{kj}^{B_\beta}} \,,\label{phi2F2_psi2Fxpsi2F}
    \\
    \left[\dot C_{\phi^2 \widetilde F^2}\right]_{ab}^{A_\alpha B_\beta} &= 8 \mathcal S_{\alpha\beta}^{-1} \Im{y^*_{jia} \left[C_{\psi^2 F}\right]_{il}^{A_\alpha} y^*_{lkb} \left[C_{\psi^2 F}\right]_{kj}^{B_\beta}} \,.\label{phi2Ftilde2_psi2Fxpsi2F}
\end{align}

\paragraph{$\phi^2F^2 \leftarrow \psi^2\phi^2 \times \phi F^2$:}

\begin{align}
    \left[\dot C_{\phi^2 F^2}\right]_{ab}^{A_\alpha B_\beta} &= -4 \Re{y_{ijc}^* \left[C_{\psi^2\phi^2}\right]_{ijab}} \left[ C_{\phi F^2}\right]_{c}^{A_\alpha B_\beta}\,,
    \\
    \left[\dot C_{\phi^2 \widetilde F^2}\right]_{ab}^{A_\alpha B_\beta} &= -4 \Re{y_{ijc}^* \left[C_{\psi^2\phi^2}\right]_{ijab}} \left[ C_{\phi \widetilde F^2}\right]_{c}^{A_\alpha B_\beta}\,.
\end{align}

\paragraph{$\phi^2F^2 \leftarrow \psi^2\phi^2 \times \psi^2 F$:}

\begin{align}
    \left[\dot C_{\phi^2 F^2}\right]_{ab}^{A_\alpha B_\beta} &= -4\mathcal{S}_{\alpha\beta}^{-1}\sum_{\sigma(\{A_\alpha,B_\beta\})} g_\alpha \Re{ t_{ij}^{A_\alpha} \left[C_{\psi^2\phi^2}\right]_{ikab}\left[C_{\psi^2 F}\right]_{kj}^{*B_\beta}}\,,
    \\
    \left[\dot C_{\phi^2 \widetilde F^2}\right]_{ab}^{A_\alpha B_\beta} &= -4 \mathcal{S}_{\alpha\beta}^{-1} \sum_{\sigma(\{A_\alpha,B_\beta\})} g_\alpha \Im{ t_{ij}^{A_\alpha} \left[C_{\psi^2\phi^2}\right]_{ikab}\left[C_{\psi^2 F}\right]_{kj}^{*B_\beta}}\,.
\end{align}

\subsubsection{\texorpdfstring{$F^3$}{F3} class}

\paragraph{$F^3 \leftarrow F^3$:}

\begin{align}
    \left[\dot C_{F^3}\right]^{A_\alpha B_\alpha C_\alpha} &=  \left.\gamma_{c,v}^{A_\alpha D_\alpha}\right|_{\text{fer}}\left[ C_{ F^3}\right]^{D_\alpha B_\alpha C_\alpha}
    + \left.\gamma_{c,v}^{B_\alpha D_\alpha}\right|_{\text{fer}}\left[ C_{F^3}\right]^{A_\alpha D_\alpha C_\alpha} \nonumber \\
    &+ \left.\gamma_{c,v}^{C_\alpha D_\alpha}\right|_{\text{fer}}\left[ C_{F^3}\right]^{A_\alpha B_\alpha D_\alpha}\,,
    \\
    \left[\dot C_{\widetilde F^3}\right]^{A_\alpha B_\alpha C_\alpha} &=  \left.\gamma_{c,v}^{A_\alpha D_\alpha}\right|_{\text{fer}}\left[ C_{ \widetilde F^3}\right]^{D_\alpha B_\alpha C_\alpha}
    + \left.\gamma_{c,v}^{B_\alpha D_\alpha}\right|_{\text{fer}}\left[ C_{\widetilde F^3}\right]^{A_\alpha D_\alpha C_\alpha} \nonumber \\
    &+ \left.\gamma_{c,v}^{C_\alpha D_\alpha}\right|_{\text{fer}}\left[ C_{\widetilde F^3}\right]^{A_\alpha B_\alpha D_\alpha}\,,
\end{align}
where $\left.\gamma^{A_\alpha B_\alpha}_{c,v}\right|_{\text{fer}} = \frac{2}{3}g_\alpha^2 \Tr(t^{A_\alpha }t^{B_\alpha})$ is the fermionic part of the gauge bosons collinear anomalous dimension in Eq.~\eqref{eq:coll_vector}.

\subsubsection{\texorpdfstring{$\psi^2 \phi^3$}{psi2phi3} class}

In addition to the results presented below, one needs to supplement the RGEs of the $\psi^2\phi^3$-class Wilson coefficients with the expressions given in Eq.~\eqref{eq:phi6<-Ctildeandbar}.
The explicit expressions for $\dot{\tildee C}_{D^2\phi^4}$ and $\dot{\overline C}_{D^2\phi^4}$ can be obtained using their definitions in Eqs.~\eqref{eq:Ctilde_def} and \eqref{eq:Cbar_def} and the RGEs for $C_{D^2\phi^4}$ reported in Sec.~\ref{sec:D2phi4RGE}.

\paragraph{\texorpdfstring{$\psi^2 \phi^3 \leftarrow \psi^2 \phi^3$}{psi2phi3frompsi2phi3}:}

\begin{align}
    \left[\dot C_{\psi^2\phi^3}\right]_{ijabc} &= 
     y_{ijd}y_{kld}^* \left[ C_{\psi^2\phi^3}\right]_{klabc}
    - \frac{1}{2!}\sum_{\sigma(\{a,b,c\})} \left(2\sum_{\alpha=1}^{N_G} g_\alpha^2 \theta^{A_\alpha}_{bd}\theta^{A_\alpha}_{ce}-\lambda_{bcde}\right)\left[ C_{\psi^2\phi^3}\right]_{ijade} \nonumber 
\\
    & + \frac{1}{2!}\sum_{\sigma(\{a,b,c\}\times \{i,j\})}\left(2 y_{jkd}y^*_{klc}+y_{jkc}y^*_{kld}+ 4 \sum_{\alpha=1}^{N_G} g_\alpha^2 t^{A_\alpha}_{lj} \theta^{A_\alpha}_{cd}\right)\left[ C_{\psi^2\phi^3}\right]_{ilabd} \nonumber \\
    &+(2y_{ild}y_{jkd}+y_{ijd}y_{kld})\left[C_{\psi^2\phi^3}\right]_{klabc}^* \nonumber
\\
    & + \frac{1}{2!}\sum_{\sigma(\{a,b,c\})} \gamma_{c,s}^{cd} \left[ C_{\psi^2\phi^3}\right]_{ijabd} + \sum_{\sigma(\{i,j\})}\gamma_{c,f}^{jk} \left[ C_{\psi^2\phi^3}\right]_{ikabc} \,.\label{psi2phi3_psi2phi3}
\end{align}
Here, $\gamma_{c,s}$ and $\gamma_{c,f}$ denote the collinear anomalous dimensions of scalar and fermion fields, respectively, as defined in Eqs.~\eqref{eq:coll_scalar} and \eqref{eq:coll_fermion}.

\paragraph{\texorpdfstring{$\psi^2 \phi^3 \leftarrow \phi^2 F^2$:}{psi2phi3fromphi2F2}}

\begin{align}
    \left[\dot C_{\psi^2\phi^3}\right]_{ijabc} &= \sum_{\sigma(\{i,j\}\times\{a,b,c\})} \sum_{\alpha=1}^{N_G}\sum_{\beta=1}^{\alpha} g_\alpha g_\beta t^{A_\alpha}_{ki} t^{B_\beta}_{lk} y_{lja} \left(\left[C_{\phi^2 F^2}\right]_{bc}^{A_\alpha B_\beta}+i\left[C_{\phi^2 \tildee F^2}\right]_{bc}^{A_\alpha B_\beta} \right) \,.
    \label{psi2phi3_phi2F2}
\end{align}

\paragraph{\texorpdfstring{$\psi^2 \phi^3 \leftarrow \psi^2 \phi F$:}{psi2phi3fromphi2phiF}}

\begin{align}
    \left[\dot C_{\psi^2\phi^3}\right]_{ijabc} &= \sum_{\sigma(\{i,j\}\times\{a,b,c\})} \sum_{\alpha=1}^{N_G} g_\alpha \left(t^{A_\alpha}_{li}y_{kma}^{*} y_{mlb} 
    + \frac{1}{2}y_{lia}^{\phantom *} y_{kld}^* \theta^{A_\alpha}_{db} 
    - \sum_{\beta=1}^{N_G} g_\beta^2 t^{B_\beta}_{ik} \theta^{B_\beta}_{da} \theta^{A_\alpha}_{db} \right)
    \nonumber \\&\times \left[C_{\psi^2\phi F}\right]_{jkc}^{A_\alpha}\,.\label{psi2phi3_psi2phiF}
\end{align}

\paragraph{\texorpdfstring{$\psi^2 \phi^3 \leftarrow \overline \psi^2 \psi^2$:}{psi2phi3frompsibar2psi2}}

\begin{align}
    \left[\dot C_{\psi^2\phi^3}\right]_{ijabc} =  \frac{4}{3}\sum_{\sigma(\{a,b,c\})} y_{knc}y_{lma}y^*_{nlb} \left[C_{\overline \psi^2 \psi^2}\right]_{kimj} -  \frac{2}{3}\lambda_{abcd}y_{kld} \left[C_{\overline \psi^2 \psi^2}\right]_{likj} \,.\label{psi2phi3_psibar2psi2}
\end{align}

\paragraph{\texorpdfstring{$\psi^2 \phi^3 \leftarrow \psi^4$:}{psi2phi3frompsi4}}

\begin{align}
    \left[\dot C_{\psi^2\phi^3}\right]_{ijabc} = -2 \sum_{\sigma(\{a,b,c\})}y_{knc}y_{kma}^*y^*_{nlb}\left[C_{\psi^4}\right]_{ijlm} + \lambda_{abcd}y^*_{kld} \left[C_{\psi^4}\right]_{ijkl}\,.
    \label{psi2phi3_psi4}
\end{align}

\paragraph{\texorpdfstring{$\psi^2 \phi^3 \leftarrow D^2\phi^4$:}{psi2phi3fromD2phi4}}

\begin{align}
    \left[\dot C_{\psi^2\phi^3}\right]_{ijabc} &= 
    \frac{1}{3}\sum_{\sigma(\{a,b,c\})}
    y_{ike}y_{jld}y^*_{klc} \left[C_{D^2\phi^4}\right]_{deab}
    \nonumber \\&
    + \frac{1}{24} \sum_{\sigma(\{a,b,c\} \times \{i,j\})} y_{ike}y_{jlc}y^*_{kld} \left(\left[C_{D^2\phi^4}\right]_{deab} + \left[C_{D^2\phi^4}\right]_{abde} 
    \right)\nonumber\\
    & +\frac{1}{3} \sum_{\sigma(\{a,b,c\} \times \{i,j\})} \sum_{\alpha=1}^{N_G} g_\alpha^2 t_{ki}^{A_\alpha}  \theta_{cd}^{A_\alpha} y_{kje} \left[C_{D^2\phi^4}\right]_{deab}\,.
    \label{psi2phi3_D2phi4}
\end{align}

\paragraph{\texorpdfstring{$\psi^2 \phi^3 \leftarrow D\overline\psi\psi\phi^2$}{psi2phi3fromDpsibarpsiphi2}:}

\begin{align}
    \left[\dot C_{\psi^2\phi^3}\right]_{ijabc} &= \sum_{\sigma(\{i,j\}\times\{a,b,c\})}  \Bigg[\sum_{\alpha=1}^{N_G} g_\alpha^2 \left(\frac{1}{4} y_{kia} t_{lj}^{A_\alpha} \theta_{bd}^{A_\alpha} \left[C_{D\overline\psi\psi\phi^2}\right]_{klcd}-\frac{1}{3} y_{lka}t_{li}^{A_\alpha} \theta_{bd}^{A_\alpha} \left[C_{D\overline\psi\psi\phi^2}\right]_{kjcd}\right.\nonumber\\
    &\left.+ \frac{1}{12} y_{kia}t_{kl}^{A_\alpha} \theta_{bd}^{A_\alpha} \left[C_{D\overline\psi\psi\phi^2}\right]_{ljcd}\right) + \frac{1}{3} y_{lka}^{\phantom *} \left( y_{mid}^{\phantom *} y_{lmb}^* +\frac{1}{2}y_{mib}^{\phantom *} y_{lmd}^* \right)  \left[C_{D\overline\psi\psi\phi^2}\right]_{kjcd}\nonumber\\
    & +\frac{1}{6} y_{kie} \lambda_{abde} \left[C_{D\overline\psi\psi\phi^2}\right]_{kjcd}\Bigg]\,.
\end{align}

\paragraph{\texorpdfstring{$\psi^2 \phi^3 \leftarrow \psi^2\phi^2\times\psi^2\phi^2$}{psi2phi3frompsi2phi2xpsi2phi2}:}

\begin{align}
    \left[\dot C_{\psi^2\phi^3}\right]_{ijabc} &= -\frac{4}{3} \sum_{\sigma(\{i,j\}\times\{a,b,c\})}
    \bigg(
    y_{ikd} \left[C_{\psi^2\phi^2}\right]_{jlad} \left[C_{\psi^2\phi^2}\right]_{klbc}^* + y_{kla}^* \left[C_{\psi^2\phi^2}\right]_{ilbd} \left[C_{\psi^2\phi^2}\right]_{jkcd}\nonumber \\
    &- 
    y_{jlc} \left[C_{\psi^2\phi^2}\right]_{ikad} \left[C_{\psi^2\phi^2}\right]_{lkbd}^*
    \bigg)\,.\label{psi2phi3_psi2phi2xpsi2phi2}
\end{align}

\paragraph{$\psi^2\phi^3 \leftarrow \phi^5 \times \psi^2\phi^2$:}

\begin{equation}
    \left[\dot C_{\psi^2\phi^3}\right]_{ijabc} = - 40 \left[C_{\phi^5}\right]_{abcde} \left[C_{\psi^2\phi^2}\right]_{ijed} \,.
\end{equation}

\paragraph{$\psi^2\phi^3 \leftarrow \psi^2\phi^2 \times \phi F^2$:}

\begin{equation}
    \left[\dot C_{\psi^2\phi^3}\right]_{ijabc} = - 2 \sum_{\sigma(\{a,b,c\})} \sum_{\alpha=1}^{N_G} \sum_{\beta=1}^{N_G} \mathcal S_{\alpha\beta}g_\alpha g_\beta t^{A_\alpha}_{ki} t^{B_\beta}_{lj}  \left[C_{\phi F^2}\right]^{A_\alpha B_\beta}_c \left[C_{\psi^2\phi^2}\right]_{klab} \,.
\end{equation}

\paragraph{$\psi^2\phi^3 \leftarrow \phi F^2 \times \phi F^2$:}

\begin{align}
    \left[\dot C_{\psi^2\phi^3}\right]_{ijabc} &= \sum_{\sigma(\{i,j\}\times\{a,b,c\})} \sum_{\alpha=1}^{N_G} \sum_{\beta=1}^{N_G}\sum_{\gamma=1}^{N_G}\frac{1}{2} \mathcal{S}_{\alpha\gamma} \mathcal{S}_{\beta\gamma}g_\alpha g_\beta t_{ki}^{A_\alpha} \left[  i t_{lk}^{B_\beta} y_{ljb} \left(\left[C_{\phi F^2}\right]_{c}^{B_\beta C_\gamma} \left[C_{\phi \tildee F^2}\right]_{a}^{A_\alpha C_\gamma}\right.\right.\nonumber\\
    &\left.\left.+ \left[C_{\phi F^2}\right]_{a}^{B_\beta C_\gamma} \left[C_{\phi \tildee F^2}\right]_{c}^{A_\alpha C_\gamma}\right) +  2 t_{lj}^{B_\beta} y_{klb} \left[C_{\phi F^2}\right]_a^{A_\alpha C_\gamma} \left[C_{\phi F^2}\right]_c^{B_\beta C_\gamma}\right]\,.
\end{align}

\paragraph{$\psi^2\phi^3 \leftarrow \psi^2 F \times \psi^2 F$:}

\begin{align}
    \left[\dot C_{\psi^2\phi^3}\right]_{ijabc} &= \sum_{\sigma(\{i,j\}\times\{a,b,c\})} \Bigg[\sum_{\alpha=1}^{N_G} \sum_{\beta=1}^{N_G} 2 g_\alpha g_\beta \theta^{A_\alpha}_{bd}\theta^{B_\beta}_{dc} y_{kla}^* \left[C_{\psi^2 F}\right]_{ki}^{A_\alpha}\left[C_{\psi^2 F}\right]_{lj}^{B_\beta}\nonumber\\
    & +\sum_{\alpha=1}^{N_G} y_{kla}^{*} y_{lmb}^{\phantom *} \, y^*_{mnc} \, \left[C_{\psi^2 F}\right]_{ki}^{A_\alpha}\left[C_{\psi^2 F}\right]_{nj}^{A_\alpha}\Bigg]\,. 
\end{align}

\paragraph{$\psi^2\phi^3 \leftarrow \phi F^2 \times \psi^2 F$:}

\begin{align}
    \left[\dot C_{\psi^2\phi^3}\right]_{ijabc} &= \sum_{\sigma(\{i,j\}\times\{a,b,c\})} \sum_{\alpha=1}^{N_G}\sum_{\beta=1}^{N_G} \mathcal{S}_{\alpha\beta} g_\alpha\Bigg[\frac{i}{2} y_{kia}^{\phantom *} t_{kl}^{A_\alpha} y_{lmb}^* \left[C_{\phi \tildee F^2}\right]_{c}^{A_\alpha B_\beta}\left[C_{\psi^2 F}\right]_{mj}^{B_\beta}\nonumber\\
    &+ t_{ki}^{A_\alpha} y_{kla}^{\phantom *} y_{lmb}^* \left[C_{\phi F^2}\right]_{c}^{A_\alpha B_\beta}\left[C_{\psi^2 F}\right]_{mj}^{B_\beta}\nonumber\\
    &+g_\beta\sum_{\gamma=1}^{N_G} g_\gamma t_{ki}^{A_\alpha}\theta_{ad}^{C_\gamma}\theta_{db}^{B_\beta}\left[C_{\phi F^2}\right]_c^{A_\alpha B_\beta} \left[C_{\psi^2 F}\right]_{kj}^{C_\gamma}\nonumber\\
    &+g_\alpha\sum_{\gamma=1}^{N_G} g_\gamma t_{ki}^{A_\alpha}\theta_{ad}^{A_\alpha}\theta_{db}^{C_\gamma}\left[C_{\phi F^2}\right]_c^{A_\alpha B_\beta} \left[C_{\psi^2 F}\right]_{kj}^{B_\beta}\Bigg]\,.
\end{align}

\paragraph{$\psi^2\phi^3 \leftarrow \psi^2\phi^2 \times \psi^2 F$:}

\begin{align}
    \left[\dot C_{\psi^2\phi^3}\right]_{ijabc} &= \sum_{\sigma(\{i,j\}\times\{a,b,c\})} \sum_{\alpha=1}^{N_G} g_\alpha \theta_{ad}^{A_\alpha} y_{kib} \left[C_{\psi^2\phi^2}\right]_{klcd}^* \left[C_{\psi^2 F}\right]_{lj}^{A_\alpha}\,.
\end{align}

\subsubsection{\texorpdfstring{$\psi^4$}{psi4} class}

\paragraph{\texorpdfstring{$\psi^4 \leftarrow \psi^4$}{psi4frompsi4}:}

\begin{align}
    \left[\dot C_{\psi^4}\right]_{ijkl} &= 
    -\sum_{\sigma(\{i,j\} \times \{k,l\})} 
    \left[ 4 \sum_{\alpha=1}^{N_G} g_\alpha^2 \left(t^{A_\alpha}_{mj}t^{A_\alpha}_{nl} - t^{A_\alpha}_{nj}t^{A_\alpha}_{ml} \right)  + y_{jla}y^*_{mna}  \right] \left[C_{\psi^4}\right]_{imkn}\nonumber \\
    &
    + y_{ija}y^*_{mna}  \left[C_{\psi^4}\right]_{klmn}
    + y_{kla}y^*_{mna}  \left[C_{\psi^4}\right]_{ijmn}
    + \sum_{\sigma(\{i,j\})} \gamma_{c,f}^{im}\left[C_{\psi^4}\right]_{mjkl}\nonumber \\
    & + \sum_{\sigma(\{k,l\})} \gamma_{c,f}^{km}\left[C_{\psi^4}\right]_{ijml}\,.
\end{align}
Here, $\gamma_{c,f}$ denotes the collinear anomalous dimension of fermion fields, as defined in Eq.~\eqref{eq:coll_fermion}.

\paragraph{\texorpdfstring{$\psi^4 \leftarrow \overline \psi^2 \psi^2$}{psi4frompsibar2psi2}:}

\begin{align}
    \left[\dot C_{\psi^4}\right]_{ijkl} &= - (2 y_{ina}y_{jma} + y_{ija}y_{mna}) \left[ C_{\overline \psi^2 \psi^2}\right]_{mknl} 
    + \binom{i \leftrightarrow k}{j \leftrightarrow l}\,.
\end{align}

\paragraph{\texorpdfstring{$\psi^4 \leftarrow \psi^2 \phi F$}{psi4frompsi2phiF}:}

\begin{align}
    \left[\dot C_{\psi^4}\right]_{ijkl} =
    \frac{1}{2}\sum_{\sigma(\{i,j\} \times \{k,l\})}\sum_{\alpha=1}^{N_G} 
    g_\alpha 
    \left(
    t^{A_\alpha}_{mi}y_{lma}-t^{A_\alpha}_{ml}y_{ima}
    \right) \left[C_{\psi^2 \phi F}\right]_{jka}^{A_\alpha}
    \,.
\end{align}

\paragraph{$\psi^4 \leftarrow \psi^2 \phi^2 \times \psi^2 \phi^2$:}

\begin{equation}
    \left[\dot C_{\psi^4}\right]_{ijkl} =
    -2 \left[C_{\psi^2\phi^2}\right]_{ijab} \left[C_{\psi^2\phi^2}\right]_{klab} \,.
\end{equation}

\paragraph{$\psi^4 \leftarrow \phi F^2 \times \phi F^2$:}

\begin{equation}
    \left[\dot C_{\psi^4}\right]_{ijkl} = \frac{1}{2}y_{ija}y_{klb} \sum_{\alpha=1}^{N_G}\sum_{\beta=1}^\alpha \mathcal S_{\alpha \beta} \left(
    \left[C_{\phi F^2}\right]^{A_\alpha B_\beta}_a
    \left[C_{\phi F^2}\right]^{A_\alpha B_\beta}_b
    +
    \left[C_{\phi \widetilde F^2}\right]^{A_\alpha B_\beta}_a
    \left[C_{\phi \widetilde F^2}\right]^{A_\alpha B_\beta}_b
    \right)\,.
\end{equation}

\paragraph{$\psi^4 \leftarrow \phi F^2 \times \psi^2 F$:}

\begin{align}
    \left[\dot C_{\psi^4}\right]_{ijkl} &= -\frac{1}{2} \sum_{\alpha=1}^{N_G} \sum_{\beta=1}^{N_G} \mathcal S_{\alpha\beta} g_\beta y_{ija}
    \left(t^{B_\beta}_{ml} \left[C_{\psi^2 F}\right]^{A_\alpha}_{mk} + t^{B_\beta}_{mk} \left[C_{\psi^2 F}\right]^{A_\alpha}_{ml} \right)
    \nonumber \\
    &\times 
    \left(\left[C_{\phi F^2}\right]^{A_\alpha B_\beta}_a - i \left[C_{\phi \widetilde F^2}\right]^{A_\alpha B_\beta}_a\right) + \binom{i \leftrightarrow k}{j \leftrightarrow l} \,.
\end{align}

\paragraph{$\psi^4 \leftarrow \psi^2 F \times \psi^2 F$:}

\begin{align}
    \left[\dot C_{\psi^4}\right]_{ijkl} &=
    -3 \sum_{\alpha=1}^{N_G} g_\alpha^2 f^{C_\alpha A_\alpha B_\alpha} f^{D_\alpha A_\alpha B_\alpha} \left(
    \left[C_{\psi^2 F}\right]_{il}^{C_\alpha} \left[C_{\psi^2 F}\right]_{jk}^{D_\alpha} + \left[C_{\psi^2 F}\right]_{ik}^{C_\alpha} \left[C_{\psi^2 F}\right]_{jl}^{D_\alpha}\right) 
    \nonumber \\
    &+ \frac{9}{2}i \sum_{\sigma(\{i,j\}\times\{k,l\})} \sum_{\alpha=1}^{N_G} g_\alpha^2 f^{C_\alpha A_\alpha B_\alpha}
    \left(
    t^{B_\alpha}_{mj} \left[C_{\psi^2 F}\right]_{ml}^{A_\alpha} - t^{B_\alpha}_{ml} \left[C_{\psi^2 F}\right]_{mj}^{A_\alpha}
    \right) \left[C_{\psi^2 F}\right]_{ik}^{C_\alpha}\nonumber \\
    &-6\sum_{\alpha=1}^{N_G}\sum_{\beta=1}^{N_G} g_\beta^2
    \left[
    \left( 
    t^{B_\beta}_{ml}t^{B_\beta}_{nk}+t^{B_\beta}_{mk}t^{B_\beta}_{nl}
    \right) \left[C_{\psi^2 F}\right]_{im}^{A_\alpha} \left[C_{\psi^2 F}\right]_{jn}^{A_\alpha}+\binom{i\leftrightarrow k}{j\leftrightarrow l}
    \right]\nonumber \\
    &+\sum_{\sigma(\{i,j\}\times\{k,l\})} \sum_{\alpha=1}^{N_G}\sum_{\beta=1}^{N_G} g_\alpha g_\beta \left(
    t^{A_\alpha}_{nl}t^{B_\beta}_{mj}-4t^{A_\alpha}_{nj}t^{B_\beta}_{ml}
    \right)
    \left[C_{\psi^2 F}\right]_{im}^{A_\alpha}
    \left[C_{\psi^2 F}\right]_{kn}^{B_\beta} \,.
\end{align}

\subsubsection{\texorpdfstring{$\psi^2 \phi F$}{psi2phiF} class}

\paragraph{$\psi^2 \phi F \leftarrow \psi^2 \phi F$:}

\begin{align}
    \left[\dot C_{\psi^2 \phi F}\right]_{ija}^{A_\alpha} &=  \sum_{\beta=1}^{N_G}\left[ 2 g_\alpha g_\beta \left(3\theta^{A_\alpha}_{ac}\theta^{B_\beta}_{cb}-2\theta^{B_\beta}_{ac}\theta^{A_\alpha}_{cb}\right)\left[ C_{\psi^2 \phi F}\right]_{ijb}^{B_\beta}+ 8 g_\beta^2 t^{B_\beta}_{li}t^{B_\beta}_{kj} \left[ C_{\psi^2 \phi F}\right]_{kla}^{A_\alpha}\right]\nonumber \\
    &+ \Bigg[\Bigg(2y_{jkb}y^*_{kla}+y_{jka}y^*_{klb}+4\sum_{\beta=1}^{N_G}g_\beta^2 t^{B_\beta}_{lj}\theta^{B_\beta}_{ab} \Bigg)\left[C_{\psi^2\phi F}\right]_{ilb}^{A_\alpha}\nonumber \\
    &- 
    4\sum_{\beta=1}^{N_G}g_\alpha g_\beta t^{A_\alpha}_{kj}t^{B_\beta}_{lk}\left[C_{\psi^2\phi F}\right]^{B_\beta}_{ila}
    -\left(i \leftrightarrow j\right)\Bigg] + \gamma_{c,f}^{ik}\left[C_{\psi^2\phi F}\right]^{A_\alpha}_{kja} \nonumber \\&+ \gamma_{c,f}^{jk}\left[C_{\psi^2\phi F}\right]^{A_\alpha}_{ika} + \gamma_{c,s}^{ab}\left[C_{\psi^2\phi F}\right]^{A_\alpha}_{ijb}
    + \gamma_{c,v}^{A_\alpha B_\alpha}\left[C_{\psi^2\phi F}\right]^{B_\alpha}_{ija}
    \,.\label{psi2phiF_psi2phiF}
\end{align}
Here, $\gamma_{c,v}$, $\gamma_{c,s}$, and $\gamma_{c,f}$ denote the collinear anomalous dimensions of gauge boson, scalar, and fermion fields, respectively, as defined in Eqs.~\eqref{eq:coll_vector}, \eqref{eq:coll_scalar}, and \eqref{eq:coll_fermion}.

\paragraph{$\psi^2 \phi F \leftarrow \phi^2 F^2$:}

\begin{equation}
    \left[\dot C_{\psi^2 \phi F}\right]_{ija}^{A_\alpha} =
    2 \sum_{\beta=1}^{N_G} \mathcal S_{\alpha\beta} g_\beta \left(t^{B_\beta}_{ki} y_{jkb}-t^{B_\beta}_{kj} y_{ikb}  \right) \left(
    \left[C_{\phi^2 F^2}\right]_{ab}^{A_\alpha B_\beta} + i \left[C_{\phi^2 \widetilde F^2}\right]_{ab}^{A_\alpha B_\beta}
    \right)\,.\label{psi2phiF_phi2F2}
\end{equation}

\paragraph{$\psi^2 \phi F \leftarrow \psi^4$:}

\begin{equation}
    \left[\dot C_{\psi^2 \phi F}\right]_{ija}^{A_\alpha} = - 4 g_\alpha \left(
    t^{A_\alpha}_{lm} y^*_{kma} - t^{A_\alpha}_{km} y^*_{lma} 
    \right)
    \left[C_{\psi^4}\right]_{ikjl}
    \,.\label{psi2phiF_psi4}
\end{equation}

\paragraph{$\psi^2 \phi F \leftarrow F^3$:}

\begin{align}
    \left[\dot C_{\psi^2 \phi F}\right]_{ija}^{A_\alpha} &=
    - 3ig_\alpha^2 \left(
    \theta^{C_\alpha}_{ab}
    t^{B_\alpha}_{ki}y_{jkb}
    - 
    t^{C_\alpha}_{ki}t^{B_\alpha}_{mk}y_{jma} - 2 t^{C_\alpha}_{ki} t^{B_\alpha}_{mj}y_{kma}
    - \left(i \leftrightarrow j \right)
    \right)\nonumber
    \\&\times 
    \left(
    \left[C_{F^3}\right]^{A_\alpha B_\alpha C_\alpha}+i \left[C_{\widetilde F^3}\right]^{A_\alpha B_\alpha C_\alpha}
    \right)\,.
    \label{psi2phiF_F3}
\end{align}

\paragraph{$\psi^2\phi F \leftarrow \phi F^2 \times \phi F^2$:}

\begin{align}
    \left[\dot C_{\psi^2 \phi F}\right]_{ija}^{A_\alpha} &= 2 \sum_{\beta=1}^{N_G} \sum_{\gamma=1}^{N_G}\mathcal S_{\alpha\beta} S_{\beta\gamma} g_\gamma t^{C_\gamma}_{ki} y_{kjb}
    \left(
    \left[C_{\phi F^2}\right]^{A_\alpha B_\beta}_b + i \left[C_{\phi \widetilde F^2}\right]^{A_\alpha B_\beta}_b
    \right)
     \left[C_{\phi F^2}\right]^{B_\beta C_\gamma}_a 
    \nonumber \\&- (i \leftrightarrow j)\,.
    \label{psi2phiF_phiF2xphiF2}
\end{align}

\paragraph{$\psi^2\phi F \leftarrow \psi^2 \phi^2 \times \phi F^2$:}

\begin{equation}
    \left[\dot C_{\psi^2 \phi F}\right]_{ija}^{A_\alpha} = -4 \sum_{\beta=1}^{N_G} \mathcal S_{\alpha\beta} g_\beta t^{B_\beta}_{kj} 
    \left(\left[C_{\phi F^2}\right]^{A_\alpha B_\beta}_b + i \left[C_{\phi \widetilde F^2}\right]^{A_\alpha B_\beta}_b \right) \left[C_{\psi^2\phi^2}\right]_{ikab}  
    - (i \leftrightarrow j)\,.
    \label{psi2phiF_phiF2xpsi2phi2}
\end{equation}

\paragraph{$\psi^2\phi F \leftarrow \psi^2\phi^2 \times \psi^2 F$:}

\begin{equation}
    \left[\dot C_{\psi^2 \phi F}\right]_{ija}^{A_\alpha} = -4 y^*_{klb} \left[C_{\psi^2\phi^2}\right]_{ikab} \left[C_{\psi^2 F}\right]_{lj}^{A_\alpha} - (i \leftrightarrow j) \,.
\end{equation}

\paragraph{$\psi^2\phi F \leftarrow \phi F^2 \times \psi^2 F$:}

\begin{align}
    \left[\dot C_{\psi^2 \phi F}\right]_{ija}^{A_\alpha} & = 2 \sum_{\beta=1}^{N_G}\sum_{\gamma=1}^{N_G} \mathcal{S}_{\beta\gamma} t_{ki}^{B_\beta} t_{lj}^{C_\gamma} \left[C_{\phi F^2}\right]_a^{B_\beta C_\gamma} \left[C_{\psi^2 F}\right]_{kl}^{A_\alpha}\nonumber\\
    &+2 g_\alpha \sum_{\beta=1}^{N_G}\sum_{\gamma=1}^{N_G} \mathcal{S}_{\beta\gamma} t_{ki}^{B_\beta} t_{lk}^{A_\alpha} \left(\left[C_{\phi F^2}\right]_a^{B_\beta C_\gamma}+i\left[C_{\phi \tildee F^2}\right]_a^{B_\beta C_\gamma}\right)\left[C_{\psi^2 F}\right]_{kl}^{A_\alpha}\nonumber\\
    &+2ig_\alpha \sum_{\beta=1}^{N_G}\mathcal{S}_{\alpha\beta} g_\beta f^{A_\alpha C_\alpha D_\alpha} t_{ki}^{B_\beta}\left(2\left[C_{\phi F^2}\right]_a^{B_\beta C_\alpha}-i\left[C_{\phi \tildee F^2}\right]_a^{B_\beta C_\alpha}\right)\left[C_{\psi^2 F}\right]_{kj}^{D_\alpha}\nonumber\\
    &+ 12 i g_\alpha^2f^{B_\alpha C_\alpha D_\alpha} t_{ki}^{B_\alpha}\left(2\left[C_{\phi F^2}\right]_a^{A_\alpha D_\alpha}+i\left[C_{\phi \tildee F^2}\right]_a^{A_\alpha D_\alpha}\right)\left[C_{\psi^2 F}\right]_{kj}^{C_\alpha}\nonumber\\
    &+ 2 i g_\alpha^2 \sum_{\beta=1}^{N_G} \mathcal{S}_{\alpha\beta} f^{A_\alpha B_\alpha C_\alpha} t_{ki}^{B_\alpha}\left(5\left[C_{\phi F^2}\right]_a^{C_\alpha D_\beta}+3i\left[C_{\phi \tildee F^2}\right]_a^{C_\alpha D_\beta}\right)\left[C_{\psi^2 F}\right]_{kj}^{D_\beta}\nonumber\\
    &+ 2 g_\alpha^2 \sum_{\beta=1}^{N_G}\sum_{\gamma=1}^{N_G} \mathcal{S}_{\beta\gamma} g_\gamma t_{ki}^{A_\alpha} t_{lj}^{C_\gamma}\left(\left[C_{\phi F^2}\right]_a^{C_\gamma B_\beta}+i\left[C_{\phi \tildee F^2}\right]_a^{C_\gamma D_\beta}\right)\left[C_{\psi^2 F}\right]_{kl}^{D_\beta}\nonumber\\
    &- 6 i \sum_{\beta=1}^{N_G}\sum_{\gamma=1}^{N_G} \mathcal{S}_{\beta\gamma} g_\beta g_\gamma t_{kl}^{B_\beta} t_{lj}^{C_\gamma}\left[C_{\phi \tildee F^2}\right]_a^{C_\gamma B_\beta}\left[C_{\psi^2 F}\right]_{ki}^{A_\alpha}\nonumber\\
    &- 8 g_\alpha^2 f^{B_\alpha C_\alpha D_\alpha} f^{B_\alpha C_\alpha E_\alpha} \left(\left[C_{\phi F^2}\right]_a^{A_\alpha D_\alpha}+i\left[C_{\phi \tildee F^2}\right]_a^{A_\alpha D_\alpha}\right)\left[C_{\psi^2 F}\right]_{ij}^{E_\alpha}\nonumber\\
    &+ 2 g_\alpha^2 f^{A_\alpha B_\alpha D_\alpha} f^{B_\alpha C_\alpha E_\alpha} \left(5\left[C_{\phi F^2}\right]_a^{C_\alpha D_\alpha}+3i\left[C_{\phi \tildee F^2}\right]_a^{C_\alpha D_\alpha}\right)\left[C_{\psi^2 F}\right]_{ij}^{E_\alpha}\nonumber\\
    &+ 4 \sum_{\beta=1}^{N_G}\Bigg[ \left(y_{kib}^{\phantom *}y_{kla}^* + \frac{1}{2}y_{kia}^{\phantom *}y_{klb}^*\right)\left(\left[C_{\phi F^2}\right]_{b}^{A_\alpha B_\beta}+i\left[C_{\phi \tildee F^2}\right]_{b}^{A_\alpha B_\beta}\right)\left[C_{\psi^2 F}\right]_{lj}^{B_\beta}\nonumber\\
    &+ \frac{1}{2}y_{kib} y_{lja}\left(\left[C_{\phi F^2}\right]_{b}^{A_\alpha B_\beta}+i\left[C_{\phi \tildee F^2}\right]_{b}^{A_\alpha B_\beta}\right)\left[C_{\psi^2 F}\right]_{lk}^{B_\beta*}\Bigg] - (i\leftrightarrow j)
    \,.
\end{align}

\paragraph{$\psi^2\phi F \leftarrow \psi^2 F \times \psi^2 F$:}

\begin{align}
    \left[\dot C_{\psi^2 \phi F}\right]_{ija}^{A_\alpha} & = 
    2i g_\alpha f^{A_\alpha B_\alpha C_\alpha}\left(
    2y^*_{kma}\left[C_{\psi^2 F}\right]_{ki}^{B_\alpha}\left[C_{\psi^2 F}\right]_{mj}^{C_\alpha}
    + 3 y_{ika}\left[C_{\psi^2 F}\right]_{kl}^{B_\alpha *}\left[C_{\psi^2 F}\right]_{lj}^{C_\alpha}
    \right)
    \nonumber \\&
    +\sum_{\beta=1}^{N_G}\Bigg[
    2 g_\alpha t^{A_\alpha}_{kl} y^*_{lma}\left[C_{\psi^2 F}\right]_{ik}^{B_\beta}\left[C_{\psi^2 F}\right]_{mj}^{B_\beta}
    + 4 g_\alpha t^{A_\alpha}_{mj} y_{ika}\left[C_{\psi^2 F}\right]_{lm}^{B_\beta}\left[C_{\psi^2 F}\right]_{kl}^{B_\beta *}
    \nonumber \\&
    + g_\beta \Bigg(
    4 t^{B_\beta}_{mi} y^*_{kla}\left[C_{\psi^2 F}\right]_{lm}^{A_\alpha}\left[C_{\psi^2 F}\right]_{kj}^{B_\beta}
    -2 t^{B_\beta}_{ki} y_{mja}\left[C_{\psi^2 F}\right]_{kl}^{A_\alpha}\left[C_{\psi^2 F}\right]_{lm}^{B_\beta *}
    \nonumber \\&
    +6 t^{B_\beta}_{ml} y_{ika}\left[C_{\psi^2 F}\right]_{mj}^{A_\alpha}\left[C_{\psi^2 F}\right]_{kl}^{B_\beta *}
    + 6 t^{B_\beta}_{lk} y_{ika}\left[C_{\psi^2 F}\right]_{mj}^{A_\alpha}\left[C_{\psi^2 F}\right]_{lm}^{B_\beta *}
    \nonumber \\&
    -6 t^{B_\beta}_{ki}y_{kla}\left[C_{\psi^2 F}\right]_{mj}^{A_\alpha}\left[C_{\psi^2 F}\right]_{lm}^{B_\beta *}
    -6 \theta^{B_\beta}_{ab}y^*_{klb} \left[C_{\psi^2 F}\right]_{lj}^{A_\alpha}\left[C_{\psi^2 F}\right]_{ik}^{B_\beta}
    \nonumber \\&
    +3 \theta^{B_\beta}_{ab}y_{ljb} \left[C_{\psi^2 F}\right]_{ik}^{A_\alpha}\left[C_{\psi^2 F}\right]_{kl}^{B_\beta *}
    \Bigg)
    \Bigg] - \left(i \leftrightarrow j\right)\,.
\end{align}

\subsubsection{\texorpdfstring{$\overline \psi^2 \psi^2$ class}{barpsi2psi2}}

\paragraph{$\overline \psi^2 \psi^2 \leftarrow \overline \psi^2 \psi^2$:}

\begin{align}
    \left[\dot C_{\overline \psi^2 \psi^2}\right]_{ijkl} &= y_{mna}y^*_{ika}\left[ C_{\overline \psi^2 \psi^2}\right]_{mjnl}
    + y_{jla}y^*_{mna}\left[ C_{\overline \psi^2 \psi^2}\right]_{imkn} \nonumber \\
    &- \frac{1}{3}\sum_{\sigma(\{i,k\}\times\{j,l\})}\Bigg[
    2 \sum_{\alpha=1}^{N_G} g_\alpha^2 (9 t^{A_\alpha}_{nl}t^{A_\alpha}_{km} + 2 t^{A_\alpha}_{kl}t^{A_\alpha}_{nm}) - 3y_{lma}y^*_{kna}
    \Bigg] \left[ C_{\overline \psi^2 \psi^2}\right]_{ijmn} \nonumber \\
    &+ \sum_{\sigma(\{i,k\})} \gamma_{c,f}^{mi} \left[ C_{\overline \psi^2 \psi^2}\right]_{mjkl} + \sum_{\sigma(\{j,l\})} \gamma_{c,f}^{jm} \left[ C_{\overline \psi^2 \psi^2}\right]_{imkl}\,.
\end{align}
Here, $\gamma_{c,f}$ denotes the collinear anomalous dimension of fermion fields, as defined in Eq.~\eqref{eq:coll_fermion}.

\paragraph{$\overline \psi^2 \psi^2 \leftarrow \psi^4$:}

\begin{align}
    \left[\dot C_{\overline \psi^2 \psi^2}\right]_{ijkl} &= 
    \frac{1}{2}(y^*_{ima}y^*_{kna} + y^*_{ina}y^*_{mka} + y^*_{ika}y^*_{mna})
    \left[C_{\psi^4}\right]_{jlmn} \nonumber \\
    &+ 
    \frac{1}{2}(y_{jma}y_{lna} + y_{jna}y_{mla} + y_{jla}y_{mna})
    \left[C_{\psi^4}\right]^*_{ikmn}
    \,.
\end{align}

\paragraph{$\overline \psi^2 \psi^2 \leftarrow D \overline \psi \psi \phi^2$:}

\begin{align}
    \left[\dot C_{\overline \psi^2 \psi^2}\right]_{ijkl} &= \frac{1}{6}\sum_{\sigma(\{i,k\}\times \{j,l\})}
    \Bigg(
    3y_{lmb}y^*_{mka}+\sum_{\alpha=1}^{N_G} g_\alpha^2 t^{A_\alpha}_{kl}\theta^{A_\alpha}_{ab}
    \Bigg)
    \left[C_{D \overline \psi \psi \phi^2}\right]_{ijab}
    \,.
\end{align}

\paragraph{$\overline \psi^2 \psi^2 \leftarrow \psi^2 \phi^2 \times \psi^2 \phi^2$:}

\begin{equation}
    \left[\dot C_{\overline \psi^2 \psi^2}\right]_{ijkl} =
    -2 \left[C_{\psi^2\phi^2}\right]_{jlab} \left[C_{\psi^2\phi^2}\right]_{ikab}^*\,.
\end{equation}

\paragraph{$\overline \psi^2 \psi^2 \leftarrow \phi F^2 \times \phi F^2$:}

\begin{align}
    \left[\dot C_{\overline \psi^2 \psi^2}\right]_{ijkl} &= \frac{1}{2}y^*_{ika}y_{jlb} \sum_{\alpha=1}^{N_G}\sum_{\beta=1}^\alpha \mathcal S_{\alpha \beta} \left(
    \left[C_{\phi F^2}\right]^{A_\alpha B_\beta}_a
    \left[C_{\phi F^2}\right]^{A_\alpha B_\beta}_b
    +
    \left[C_{\phi \widetilde F^2}\right]^{A_\alpha B_\beta}_a
    \left[C_{\phi \widetilde F^2}\right]^{A_\alpha B_\beta}_b
    \right) \nonumber \\
    &+ \frac{1}{3}\sum_{\alpha=1}^{N_G}\sum_{\beta=1}^{N_G}\sum_{\gamma=1}^{N_G}
    \mathcal S_{\alpha\beta}\mathcal S_{\beta\gamma} g_\alpha g_\gamma \left(t^{A_\alpha}_{ij} t^{C_\gamma}_{kl} + t^{A_\alpha}_{il} t^{C_\gamma}_{kj}\right)\nonumber \\
    &\times
    \left(
    \left[C_{\phi F^2}\right]^{A_\alpha B_\beta}_a
    \left[C_{\phi F^2}\right]^{ B_\beta C_\gamma}_a
    +
    \left[C_{\phi \widetilde F^2}\right]^{A_\alpha B_\beta}_a
    \left[C_{\phi \widetilde F^2}\right]^{ B_\beta C_\gamma}_a
    \right)\,.
\end{align}

\paragraph{$\overline \psi^2 \psi^2 \leftarrow \phi F^2 \times \psi^2 F$:}

\begin{align}
    \left[\dot C_{\overline \psi^2 \psi^2}\right]_{ijkl} &= \frac{1}{2} \sum_{\sigma(\{i,k\} \times \{j,l\})} \sum_{\alpha = 1}^{N_G} \sum_{\beta = 1}^{N_G} \mathcal{S}_{\alpha\beta} g_\beta t_{ij}^{B_\beta} \left(y_{mla} \left[C_{\psi^2 F}\right]^{A_\alpha *}_{km} + y^*_{mka} \left[C_{\psi^2 F}\right]^{A_\alpha }_{lm} \right)
    \nonumber \\
    &\times\left[C_{\phi F^2}\right]_{a}^{A_\alpha B_\beta}\,.
\end{align}

\paragraph{$\overline \psi^2 \psi^2 \leftarrow \psi^2 F\times \psi^2 F$:}

\begin{align}
    \left[\dot C_{\overline \psi^2 \psi^2}\right]_{ijkl} &=
    \sum_{\sigma(\{i,k\} \times \{j,l\})}\sum_{\alpha=1}^{N_G}
    \Bigg[
 \frac{5}{6}i g_\alpha^2 f^{A_\alpha B_\alpha C_\alpha} t^{C_\alpha}_{kl} \left[C_{\psi^2 F}\right]_{mj}^{B_\alpha}\left[C_{\psi^2 F}\right]_{im}^{A_\alpha *}
 \nonumber \\&
 +\sum_{\beta=1}^{N_G}\Bigg( \frac{1}{3} g_\alpha g_\beta t^{A_\alpha}_{ij} t^{B_\beta}_{kl}\left[C_{\psi^2 F}\right]_{mn}^{A_\alpha}\left[C_{\psi^2 F}\right]_{nm}^{B_\beta *} 
+ 2  g_\alpha g_\beta t^{A_\alpha}_{nj} t^{B_\beta}_{kl} \left[C_{\psi^2 F}\right]_{im}^{A_\alpha *}\left[C_{\psi^2 F}\right]_{mn}^{B_\beta} 
\nonumber \\&
+ 2  g_\alpha g_\beta t^{A_\alpha}_{in} t^{B_\beta }_{kl} \left[C_{\psi^2 F}\right]_{jm}^{A_\alpha}\left[C_{\psi^2 F}\right]_{mn}^{B_\beta *} 
+ 3  g_\beta^2 t^{B_\beta}_{km} t^{B_\beta}_{nl} \left[C_{\psi^2 F}\right]_{jn}^{A_\alpha}\left[C_{\psi^2 F}\right]_{im}^{A_\alpha *} 
\nonumber \\&
+3  g_\alpha g_\beta t^{A_\alpha}_{nj} t^{B_\beta}_{km} \left[C_{\psi^2 F}\right]_{nl}^{B_\beta}\left[C_{\psi^2 F}\right]_{im}^{A_\alpha *} 
+ \frac{9}{2}  g_\alpha g_\beta t^{A_\alpha}_{kn} t^{B_\beta}_{nl} \left[C_{\psi^2 F}\right]_{mj}^{B_\beta}\left[C_{\psi^2 F}\right]_{im}^{A_\alpha *} \Bigg)
\nonumber \\&
+ \frac{3}{4}y_{mja}y_{nla}\left[C_{\psi^2 F}\right]^{A_\alpha *}_{im}\left[C_{\psi^2 F}\right]^{A_\alpha *}_{kn}
+ \frac{3}{2}y_{mja}y_{kna}^*\left[C_{\psi^2 F}\right]^{A_\alpha}_{nl}\left[C_{\psi^2 F}\right]^{A_\alpha *}_{im}
\nonumber \\&
+ \frac{3}{4}y_{ima}^*y_{kna}^*\left[C_{\psi^2 F}\right]^{A_\alpha}_{mj}\left[C_{\psi^2 F}\right]^{A_\alpha}_{nl}
    \Bigg]\,.
\end{align}

\subsubsection{\texorpdfstring{$D\overline \psi \psi \phi^2$ class}{Dbarpsipsiphi2}}

\paragraph{$D\overline \psi \psi \phi^2 \leftarrow D\overline \psi \psi \phi^2$:}

\begin{align}
    \left[\dot C_{D\overline \psi \psi \phi^2}\right]_{ijab} &=
    \frac{1}{3}\left[
    2 \sum_{\alpha=1}^{N_G} g_\alpha^2 \left(9 t^{A_\alpha}_{lj} t^{A_\alpha}_{ik} + 2 t^{A_\alpha}_{ij} t^{A_\alpha}_{lk} \right)
    - 3 y_{jkc}y^*_{ilc}
    \right] \left[C_{D\overline \psi \psi \phi^2}\right]_{klab}\nonumber \\
    &- \frac{1}{3} \sum_{\alpha=1}^{N_G}g_\alpha^2 \left(
    \theta^{A_\alpha}_{ab} \theta^{A_\alpha}_{cd}
    +24 \theta^{A_\alpha}_{ac} \theta^{A_\alpha}_{bd}
    \right) \left[C_{D\overline \psi \psi \phi^2}\right]_{ijcd} \nonumber \\
    &- \left[\left(
    y_{klc}y^*_{lia}+2y_{kla}y^*_{lic}+4\sum_{\alpha=1}^{N_G}g_\alpha^2 t^{A_\alpha}_{ik}\theta^{A_\alpha}_{ca}
    \right)
    \left[C_{D\overline \psi \psi \phi^2}\right]_{kjcb}
    -(a\leftrightarrow b)\right] \nonumber \\
    &+  
    \left[\left(
    2y_{jlc}y^*_{lka}+y_{jla}y^*_{lkc}-4\sum_{\alpha=1}^{N_G}g_\alpha^2 t^{A_\alpha}_{kj}\theta^{A_\alpha}_{ca}
    \right)
    \left[C_{D\overline \psi \psi \phi^2}\right]_{ikcb}
    -(a\leftrightarrow b)\right]\nonumber \\
    &+ 
    \gamma_{c,f}^{ki} \left[C_{D\overline \psi \psi \phi^2}\right]_{kjab} + \gamma_{c,f}^{jk} \left[C_{D\overline \psi \psi \phi^2}\right]_{ikab}
    + \gamma_{c,s}^{ac}\left[C_{D\overline \psi \psi \phi^2}\right]_{ijcb} \nonumber \\
    &+ \gamma_{c,s}^{bc}\left[C_{D\overline \psi \psi \phi^2}\right]_{ijac}\,.
\end{align}
Here, $\gamma_{c,s}$ and $\gamma_{c,f}$ denote the collinear anomalous dimensions of scalar and fermion fields, respectively, as defined in Eqs.~\eqref{eq:coll_scalar} and \eqref{eq:coll_fermion}.

\paragraph{$D\overline \psi \psi \phi^2 \leftarrow \overline \psi^2 \psi^2$:}

\begin{align}
    \left[\dot C_{D\overline \psi \psi \phi^2}\right]_{ijab} =
    -\frac{4}{3}\left(
    3y_{kmb}y^*_{mla}-3y_{kma}y^*_{mlb}
    + 2 \sum_{\alpha=1}^{N_G}g_\alpha^2 t^{A_\alpha}_{lk}\theta^{A_\alpha}_{ab}
    \right)\left[C_{\overline\psi^2\psi^2}\right]_{ijkl}
    \,.
\end{align}

\paragraph{$D\overline \psi \psi \phi^2 \leftarrow D^2\phi^4$:}

\begin{align}
    \left[\dot C_{D\overline \psi \psi \phi^2}\right]_{ijab} =
    -\frac{1}{6}
    \left(
    3 y_{jkd}y^*_{kic} +\sum_{\alpha=1}^{N_G} g_\alpha^2 t^{A_\alpha}_{ij}\theta^{A_\alpha}_{cd}
    \right)
    \left[\widehat C_{D^2\phi^4}\right]_{acbd}
    \,.
\end{align}

\paragraph{$D\overline \psi \psi \phi^2 \leftarrow \psi^2\phi^2 \times \psi^2\phi^2$:}

\begin{equation}
     \left[\dot C_{D\overline \psi \psi \phi^2}\right]_{ijab} =
     -4 \left(
     \left[C_{\psi^2\phi^2}\right]_{jkbc} \left[C_{\psi^2\phi^2}\right]_{ikac}^* -  \left[C_{\psi^2\phi^2}\right]_{jkac} \left[C_{\psi^2\phi^2}\right]_{ikbc}^* \right) \,.
\end{equation}

\paragraph{$D\overline \psi \psi \phi^2 \leftarrow \phi F^2 \times \phi F^2$:}

\begin{align}
    \left[\dot C_{D\overline \psi \psi \phi^2}\right]_{ijab} &= 3 \sum_{\alpha=1}^{N_G} \sum_{\beta=1}^{N_G} \sum_{\gamma=1}^{N_G} \mathcal S_{\alpha\gamma} \mathcal S_{\beta\gamma} g_\alpha g_\beta 
    \left(t^{A_\alpha}_{kj} t^{B_\beta}_{ik} + t^{A_\alpha}_{ik} t^{B_\beta}_{kj} \right)
    \nonumber \\
    &\times
    \left(\left[C_{\phi F^2}\right]_b^{B_\beta C_\gamma} \left[C_{\phi \widetilde F^2}\right]_a^{A_\alpha C_\gamma} - \left[C_{\phi F^2}\right]_a^{B_\beta C_\gamma} \left[C_{\phi \widetilde F^2}\right]_b^{A_\alpha C_\gamma} \right) \nonumber \\
    & - \frac{2}{3} \sum_{\alpha=1}^{N_G} \sum_{\beta=1}^{N_G} \sum_{\gamma=1}^{N_G} \mathcal S_{\alpha\gamma} \mathcal S_{\beta\gamma} g_\alpha g_\beta t^{A_\alpha}_{ij}\theta^{B_\beta}_{ab}
    \nonumber \\
    &\times
    \left(\left[C_{\phi F^2}\right]_c^{B_\beta C_\gamma} \left[C_{\phi F^2}\right]_c^{A_\alpha C_\gamma} + \left[C_{\phi \widetilde F^2}\right]_c^{B_\beta C_\gamma} \left[C_{\phi \widetilde F^2}\right]_c^{A_\alpha C_\gamma} \right)\,.
\end{align}

\paragraph{$D\overline \psi \psi \phi^2 \leftarrow \psi^2\phi^2 \times \psi^2 F$:}

\begin{align}
    \left[\dot C_{D\overline \psi \psi \phi^2}\right]_{ijab} &=-12 \sum_{\alpha=1}^{N_G} g_\alpha \theta^{A_\alpha}_{bc}
    \left(
    \left[C_{\psi^2 F}\right]_{ik}^{A_\alpha *} \left[C_{\psi^2\phi^2}\right]_{kjac} + \left[C_{\psi^2 F}\right]_{jk}^{A_\alpha} \left[C_{\psi^2\phi^2}\right]^*_{kiac}
    \right)\nonumber \\& - (a \leftrightarrow b)\,.
\end{align}

\paragraph{$D\overline \psi \psi \phi^2 \leftarrow \phi F^2 \times \psi^2 F$:}

\begin{align}
    \left[\dot C_{D\overline \psi \psi \phi^2}\right]_{ijab} &= -2\sum_{\alpha=1}^{N_G} \sum_{\beta=1}^{N_G} \mathcal S_{\alpha\beta} g_\alpha \theta^{A_\alpha}_{ab}
    \Bigg[
    y_{jkc}
    \left[C_{\psi^2 F}\right]^{B_\beta *}_{ki}
    \left(
    \left[C_{\phi F^2}\right]^{A_\alpha B_\beta}_c + i \left[C_{\phi \widetilde F^2}\right]^{A_\alpha B_\beta}_c
    \right)
    \nonumber \\
    & + 
    y_{ikc}^*
    \left[C_{\psi^2 F}\right]^{B_\beta }_{kj}
    \left(
    \left[C_{\phi F^2}\right]^{A_\alpha B_\beta}_c - i \left[C_{\phi \widetilde F^2}\right]^{A_\alpha B_\beta}_c
    \right)
    \Bigg]
    \nonumber \\
    &+ 
    \sum_{\alpha=1}^{N_G} \sum_{\beta=1}^{N_G} \mathcal S_{\alpha\beta} g_\beta
    \Bigg[
    6i\left(t^{A_\alpha}_{kj} y_{klb} \left[C_{\psi^2 F}\right]^{B_\beta *}_{li} + t^{A_\alpha}_{ik} y_{klb}^* \left[C_{\psi^2 F}\right]^{B_\beta}_{lj}  \right)
    \left[C_{\phi \widetilde F^2}\right]^{A_\alpha B_\beta}_{a}
    \nonumber \\
    & + 4 
    t^{A_\alpha}_{ik}y_{ljb}\left[C_{\psi^2 F}\right]^{B_\beta *}_{kl}
    \left(
    \left[C_{\phi F^2}\right]^{A_\alpha B_\beta}_a + i \left[C_{\phi \widetilde F^2}\right]^{A_\alpha B_\beta}_a
    \right)
    \nonumber \\
    & - 4 
    t^{A_\alpha}_{kj}y^*_{lib}\left[C_{\psi^2 F}\right]^{B_\beta}_{kl}
    \left(
    \left[C_{\phi F^2}\right]^{A_\alpha B_\beta}_a - i \left[C_{\phi \widetilde F^2}\right]^{A_\alpha B_\beta}_a
    \right)
    \nonumber \\
    & + 2 
    t^{A_\alpha}_{lk}y_{jkb}\left[C_{\psi^2 F}\right]^{B_\beta *}_{li}
    \left(2
    \left[C_{\phi F^2}\right]^{A_\alpha B_\beta}_a - i \left[C_{\phi \widetilde F^2}\right]^{A_\alpha B_\beta}_a
    \right)
    \nonumber \\
    & - 2 
    t^{A_\alpha}_{kl}y^*_{ikb}\left[C_{\psi^2 F}\right]^{B_\beta}_{lj}
    \left(2
    \left[C_{\phi F^2}\right]^{A_\alpha B_\beta}_a + i \left[C_{\phi \widetilde F^2}\right]^{A_\alpha B_\beta}_a
    \right)-\left(a\leftrightarrow b\right)
    \Bigg]\,.
\end{align}

\paragraph{$D \overline \psi \psi \phi^2 \leftarrow \psi^2 F \times \psi^2 F$:}

\begin{align}
    \left[\dot C_{D\overline \psi \psi \phi^2}\right]_{ijab} &= 
    \sum_{\alpha=1}^{N_G} \sum_{\beta=1}^{N_G} 
    \Bigg[
    \frac{16}{3}g_\alpha^2 t^{A_\alpha}_{kl}\theta^{A_\alpha}_{ab}
    \left[C_{\psi^2 F}\right]_{lj}^{B_\beta} \left[C_{\psi^2 F}\right]_{ik}^{B_\beta *} 
    \nonumber\\
    &- \frac{4}{3}g_\alpha g_\beta t^{B_\beta}_{ij}\theta^{A_\alpha}_{ab} \left(
    \left[C_{\psi^2 F}\right]_{kl}^{B_\beta} \left[C_{\psi^2 F}\right]_{lk}^{A_\alpha *} + \left[C_{\psi^2 F}\right]_{kl}^{A_\alpha} \left[C_{\psi^2 F}\right]_{lk}^{B_\beta *} 
    \right) \nonumber \\
    &- 8 g_{\alpha} g_\beta \theta^{A_\alpha}_{ab}\left(
    t^{B_\beta}_{il}
    \left[C_{\psi^2 F}\right]_{jk}^{B_\beta}
    \left[C_{\psi^2 F}\right]_{kl}^{A_\alpha *}
    +
    t^{B_\beta}_{lj}
    \left[C_{\psi^2 F}\right]_{kl}^{A_\alpha}
    \left[C_{\psi^2 F}\right]_{ik}^{B_\beta *}
    \right)
    \Bigg] \nonumber \\
    &+\frac{10}{3}i\sum_{\alpha=1}^{N_G}  g_\alpha^2 f^{A_\alpha B_\alpha C_\alpha}\theta^{C_\alpha}_{ab}\left[C_{\psi^2 F}\right]_{ki}^{A_\alpha}
    \left[C_{\psi^2 F}\right]_{ik}^{B_\alpha *} \nonumber \\
    &+ \left(4\sum_{\alpha=1}^{N_G}
    y_{mjb}y^*_{ika} \left[C_{\psi^2 F}\right]_{kl}^{A_\alpha} \left[C_{\psi^2 F}\right]_{lm}^{A_\alpha *} - \left(a \leftrightarrow b\right)
    \right)\,.
\end{align}

\subsection{Running of dimension-5 operators}

\subsubsection{\texorpdfstring{$\phi^5$}{phi5} class}

In addition to the results presented below, one needs to supplement the RGEs of the $\phi^5$-class Wilson coefficients with the expressions given in Eq.~\eqref{eq:phi6<-Ctildeandbar}.
The explicit expressions for $\dot{\tildee C}_{D^2\phi^4}$ and $\dot{\overline C}_{D^2\phi^4}$ can be obtained using their definitions in Eqs.~\eqref{eq:Ctilde_def} and \eqref{eq:Cbar_def} and the RGEs for $C_{D^2\phi^4}$ reported in Sec.~\ref{sec:D2phi4RGE}.

\paragraph{$\phi^5 \leftarrow \phi^5$:}

\begin{equation}
    \left[\dot C_{\phi^5}\right]_{abcde} = \frac{1}{4!} \sum_{\sigma(\{a,b,c,d,e\})} \left.\gamma_{c,s}^{ef} \right|_{\text{fer}}\left[C_{\phi^5}\right]_{abcdf} \,,
\end{equation}
where $\left.\gamma_{c,s}^{ab} \right|_{\text{fer}} = \frac{1}{2}(y_{ija} y_{ijb}^* + y_{ijb} y_{ija}^*)$ is the fermionic part of the collinear anomalous dimension of scalar fields in Eq.~\eqref{eq:coll_scalar}.

\paragraph{\texorpdfstring{$\phi^5\leftarrow \psi^2\phi^3$}{phi5frompsi2phi3}:}

\begin{align}
    \left[\dot{C}_{\phi^5}\right]_{abcde} &= - \frac{8}{5!}\sum_{\sigma(\{a,b,c,d,e\})} \Re{\left(m_{ij}\, y_{jka}^*+y_{ija}m_{jk}^*\right) y_{klb}\left[C_{\psi^2\phi^3}\right]_{licde}^*}\,.\label{phi5_psi2phi3}
\end{align}

\paragraph{\texorpdfstring{$\phi^5\leftarrow \psi^2\phi^2$}{phi5frompsi2phi2}:}

\begin{align}
    \left[\dot{C}_{\phi^5}\right]_{abcde} &= - \frac{8}{5!}\sum_{\sigma(\{a,b,c,d,e\})} \Re{y_{ija}\, y_{jkb}^*\, y_{klc}\left[C_{\psi^2\phi^2}\right]_{lide}^*}\nonumber\\
    & +\frac{1}{6}\frac{1}{5!}\sum_{\sigma(\{a,b,c,d,e\})} \Re{y_{ijf}\lambda_{abcf}\left[C_{\psi^2\phi^2}\right]_{jide}^*}\,.\label{phi5_psi2phi2}
\end{align}

\paragraph{\texorpdfstring{$\phi^5\leftarrow \psi^2\phi^2\times\psi^2\phi^2$}{phi5frompsi2phi2xpsi2phi2}:}

\begin{align}
    \left[\dot C_{\phi^5}\right]_{abcde} &= \frac{16}{5!} \sum_{\sigma(\{a,b,c,d,e\})}\Bigg[\Re{\left[C_{\psi^2\phi^2}\right]_{ijab} y^*_{jkc} \left[C_{\psi^2\phi^2}\right]_{klde} m^*_{li}}\label{phi5_psi2phi2xpsi2phi2}\\\nonumber
     &+ \left[C_{\psi^2\phi^2}\right]_{ijab}\left(m^*_{jk} y_{klc} + y_{jkc}^* m_{lk}\right) \left[C_{\psi^2\phi^2}\right]^*_{lide}\Bigg]\,.
\end{align}

\subsubsection{\texorpdfstring{$\phi F^2$}{phiF2} class}

\paragraph{$\phi F^2 \leftarrow \phi F^2$:}

\begin{align}
    \left[\dot{C}_{\phi F^2}\right]_{a}^{A_\alpha B_\beta} &= \left.\gamma_{c,s}^{ab}\right|_{\text{fer}}\left[ C_{\phi  F^2}\right]_{b}^{A_\alpha B_\beta} + \left.\gamma^{A_\alpha C_\alpha}_{c,v}\right|_{\text{fer}}\left[ C_{\phi  F^2}\right]_{a}^{C_\alpha B_\beta} 
    + \left.\gamma^{B_\beta C_\beta}_{c,v}\right|_{\text{fer}}\left[ C_{\phi  F^2}\right]_{a}^{A_\alpha C_\beta} \,,
    \\
    \left[\dot{C}_{\phi \widetilde  F^2}\right]_{a}^{A_\alpha B_\beta} &= \left.\gamma_{c,s}^{ab}\right|_{\text{fer}}\left[ C_{\phi  \widetilde F^2}\right]_{b}^{A_\alpha B_\beta} + \left.\gamma^{A_\alpha C_\alpha}_{c,v}\right|_{\text{fer}}\left[ C_{\phi \widetilde  F^2}\right]_{a}^{C_\alpha B_\beta} 
    + \left.\gamma^{B_\beta C_\beta}_{c,v}\right|_{\text{fer}}\left[ C_{\phi \widetilde  F^2}\right]_{a}^{A_\alpha C_\beta} \,,
\end{align}
where $\left.\gamma_{c,s}^{ab} \right|_{\text{fer}} = \frac{1}{2}(y_{ija} y_{ijb}^* + y_{ijb} y_{ija}^*)$ is the fermionic part of the collinear anomalous dimension of scalar fields in Eq.~\eqref{eq:coll_scalar} and $\left.\gamma^{A_\alpha B_\alpha}_{c,v}\right|_{\text{fer}} = \frac{2}{3}g_\alpha^2 \Tr(t^{A_\alpha }t^{B_\alpha})$ the fermionic part of the gauge bosons collinear anomalous dimension in Eq.~\eqref{eq:coll_vector}.

\paragraph{$\phi F^2 \leftarrow \psi^2 \phi F$:}

\begin{align}
    \left[\dot C_{\phi F^2}\right]_{a}^{A_\alpha B_\beta} &= 4 \mathcal S_{\alpha\beta}^{-1} \sum_{\sigma(\{A_\alpha,B_\beta\})}
     g_\beta \Re{
     m^*_{ik}t^{B_\beta}_{jk}
    \left[C_{\psi^2\phi F}\right]_{ija}^{A_\alpha}
    }
    \,,\label{phiF2_psi2phiF}
    \\
    \left[\dot C_{\phi \widetilde F^2}\right]_{a}^{A_\alpha B_\beta} &= 4 \mathcal S_{\alpha\beta}^{-1} \sum_{\sigma(\{A_\alpha,B_\beta\})}
     g_\beta \Im{
     m^*_{ik}t^{B_\beta}_{jk}
    \left[C_{\psi^2\phi F}\right]_{ija}^{A_\alpha}
    }
    \,.\label{phiFtilde2_psi2phiF}
\end{align}

\paragraph{$\phi F^2 \leftarrow \psi^2 F$:}

\begin{align}
    \left[\dot C_{\phi F^2}\right]_{a}^{A_\alpha B_\beta} &= 4 \mathcal S_{\alpha\beta}^{-1} \sum_{\sigma(\{A_\alpha,B_\beta\})}
     g_\beta \Re{
     y^*_{ika}t^{B_\beta}_{jk}
    \left[C_{\psi^2 F}\right]_{ij}^{A_\alpha}
    }
    \,,\label{phiF2_psi2F}
    \\
    \left[\dot C_{\phi \widetilde F^2}\right]_{a}^{A_\alpha B_\beta} &= 4 \mathcal S_{\alpha\beta}^{-1} \sum_{\sigma(\{A_\alpha,B_\beta\})}
     g_\beta \Im{
     y^*_{ika}t^{B_\beta}_{jk}
    \left[C_{\psi^2 F}\right]_{ij}^{A_\alpha}
    }
    \,.\label{phiFtilde2_psi2F}
\end{align}

\paragraph{$\phi F^2\leftarrow \psi^2 F \times \psi^2 F$:}

\begin{align}
    \left[\dot C_{\phi F^2}\right]_{a}^{A_\alpha B_\beta} &= 32 \mathcal S_{\alpha\beta}^{-1} \Re{m^*_{ji} \left[C_{\psi^2 F}\right]_{il}^{A_\alpha} y^*_{lka} \left[C_{\psi^2 F}\right]_{kj}^{B_\beta}} \,,\label{phiF2_psi2Fxpsi2F}
    \\
    \left[\dot C_{\phi \widetilde F^2}\right]_{a}^{A_\alpha B_\beta} &= 32 \mathcal S_{\alpha\beta}^{-1} \Im{m^*_{ji} \left[C_{\psi^2 F}\right]_{il}^{A_\alpha} y^*_{lka} \left[C_{\psi^2 F}\right]_{kj}^{B_\beta}} \,.\label{phiFtilde2_psi2Fxpsi2F}
\end{align}

\paragraph{$\phi F^2 \leftarrow \phi F^2 \times \psi^2 F$:}

\begin{align}
    \left[\dot C_{\phi F^2}\right]_{a}^{A_\alpha B_\beta} &= 4g_\alpha \Re{t^{A_\alpha}_{ij}m^*_{jk}\left[C_{\psi^2 F}\right]^{C_\alpha}_{ki} + t^{C_\alpha}_{ij}m^*_{jk}\left[C_{\psi^2 F}\right]^{A_\alpha}_{ki} }\left[C_{\phi F^2}\right]_a^{C_\alpha B_\beta} \nonumber \\
    &+ 
    4g_\beta \Re{t^{B_\beta}_{ij}m^*_{jk}\left[C_{\psi^2 F}\right]^{C_\beta}_{ki} + t^{C_\beta}_{ij}m^*_{jk}\left[C_{\psi^2 F}\right]^{B_\beta}_{ki} }\left[C_{\phi F^2}\right]_a^{A_\alpha C_\beta}
    \,,
    \\
    \left[\dot C_{\phi \tildee F^2}\right]_{a}^{A_\alpha B_\beta} &= 4g_\alpha \Re{t^{A_\alpha}_{ij}m^*_{jk}\left[C_{\psi^2 F}\right]^{C_\alpha}_{ki} + t^{C_\alpha}_{ij}m^*_{jk}\left[C_{\psi^2 F}\right]^{A_\alpha}_{ki} }\left[C_{\phi \tildee F^2}\right]_a^{C_\alpha B_\beta} \nonumber \\
    &+ 
    4g_\beta \Re{t^{B_\beta}_{ij}m^*_{jk}\left[C_{\psi^2 F}\right]^{C_\beta}_{ki} + t^{C_\beta}_{ij}m^*_{jk}\left[C_{\psi^2 F}\right]^{B_\beta}_{ki} }\left[C_{\phi \tildee F^2}\right]_a^{A_\alpha C_\beta}\,.
\end{align}

\subsubsection{\texorpdfstring{$\psi^2 \phi^2$}{psi2phi2} class}

\paragraph{\texorpdfstring{$\psi^2 \phi^2 \leftarrow \psi^2 \phi^3$}{psi2phi2frompsi2phi3}:}

\begin{align}
    \left[\dot C_{\psi^2\phi^2}\right]_{ijab} = 3\sum_{\sigma(\{a,b\})} h_{acd}\left[ C_{\psi^2\phi^3}\right]_{ijbcd} 
+ \frac{3}{2}\sum_{\sigma(\{a,b\}\times \{i,j\})}\left(2 y_{jkc}m^*_{kl}+m_{jk}y^*_{klc}\right)\left[ C_{\psi^2\phi^3}\right]_{ilabc}\,.\label{psi2phi2_psi2phi3}
\end{align}

\paragraph{\texorpdfstring{$\psi^2 \phi^2 \leftarrow \psi^2 \phi^2$}{psi2phi2frompsi2phi2}:}

\begin{align}
    \left[\dot C_{\psi^2\phi^2}\right]_{ijab} &= 
    y_{ijd}y_{kld}^* \left[ C_{\psi^2\phi^2}\right]_{klab}
    - \frac{1}{2!}\sum_{\sigma(\{a,b\})} \left(2\sum_{\alpha=1}^{N_G} g_\alpha^2 \theta^{A_\alpha}_{ac}\theta^{A_\alpha}_{bd}-\lambda_{abcd}\right)\left[ C_{\psi^2\phi^2}\right]_{ijcd} \nonumber 
\\
    & + \sum_{\sigma(\{a,b\}\times \{i,j\})}\left(2 y_{jkc}y^*_{klb}+y_{jkb}y^*_{klc} + 4 \sum_{\alpha=1}^{N_G} g_\alpha^2 t^{A_\alpha}_{lj} \theta^{A_\alpha}_{bc}\right)\left[ C_{\psi^2\phi^2}\right]_{ilac} \nonumber \\
    &+\sum_{\sigma(\{a,b\}\times \{i,j\})}\Bigg( y_{jla} y_{ikc} \left[C_{\psi^2\phi^2}\right]_{lkbc} \Bigg) +\Bigg(2y_{ild}y_{jkd}+y_{ijd}y_{kld}\Bigg)\left[C_{\psi^2\phi^2}\right]_{klab}^* \nonumber\\
    & + \sum_{\sigma(\{a,b\})} \gamma_{c,s}^{bd} \left[ C_{\psi^2\phi^2}\right]_{ijad} + \sum_{\sigma(\{i,j\})}\gamma_{c,f}^{jk} \left[ C_{\psi^2\phi^2}\right]_{ikab} \,.\label{psi2phi3_psi2phi2}
\end{align}
Here, $\gamma_{c,s}$ and $\gamma_{c,f}$ denote the collinear anomalous dimensions of scalar and fermion fields, respectively, as defined in Eqs.~\eqref{eq:coll_scalar} and \eqref{eq:coll_fermion}.

\paragraph{\texorpdfstring{$\psi^2 \phi^2 \leftarrow \phi^2 F^2$:}{psi2phi2fromphi2F2}}

\begin{align}
    \left[\dot C_{\psi^2\phi^2}\right]_{ijab} &= 3\sum_{\sigma(\{i,j\}\times\{a,b\})} \sum_{\alpha=1}^{N_G}\sum_{\beta=1}^{\alpha} g_\alpha g_\beta t^{A_\alpha}_{ki} t^{B_\beta}_{lk} m_{lj} \left(\left[C_{\phi^2 F^2}\right]_{ab}^{A_\alpha B_\beta}+i\left[C_{\phi^2 \tildee F^2}\right]_{ab}^{A_\alpha B_\beta} \right) \,.
    \label{psi2phi2_phi2F2}
\end{align}

\paragraph{\texorpdfstring{$\psi^2 \phi^2 \leftarrow \phi F^2$:}{psi2phi2fromphiF2}}

\begin{align}
    \left[\dot C_{\psi^2\phi^2}\right]_{ijab} &= 3\sum_{\sigma(\{i,j\}\times\{a,b\})} \sum_{\alpha=1}^{N_G}\sum_{\beta=1}^{\alpha} g_\alpha g_\beta\Bigg[ t^{A_\alpha}_{ki} t^{B_\beta}_{lk} y_{lja} \left(\left[C_{\phi F^2}\right]_{b}^{A_\alpha B_\beta}+i\left[C_{\phi \tildee F^2}\right]_{b}^{A_\alpha B_\beta} \right)\nonumber\\
    &+t_{ki}^{A_\alpha} t_{lj}^{B_\beta} y_{kla}\left[C_{\phi F^2}\right]^{A_\alpha B_\beta}_b\Bigg]\,.
    \label{psi2phi2_phiF2}
\end{align}

\paragraph{\texorpdfstring{$\psi^2 \phi^2 \leftarrow \psi^2 \phi F$:}{psi2phi2fromphi2phiF}}

\begin{align}
    \left[\dot C_{\psi^2\phi^2}\right]_{ijab} &= 3\sum_{\sigma(\{i,j\}\times\{a,b\})} \sum_{\alpha=1}^{N_G} g_\alpha \left(t^{A_\alpha}_{li}\left(m_{km}^{*} y_{mla}+y_{kma}^{*} m_{ml}\right) 
    + \frac{1}{2}m_{li}^{\phantom *} y_{kld}^* \theta^{A_\alpha}_{da}\right)\left[C_{\psi^2\phi F}\right]_{jkb}^{A_\alpha}\,.\label{psi2phi2_psi2phiF}
\end{align}

\paragraph{\texorpdfstring{$\psi^2 \phi^2 \leftarrow \psi^2 F$:}{psi2phi2fromphi2F}}

\begin{align}
    \left[\dot C_{\psi^2\phi^2}\right]_{ijab} &= 3\sum_{\sigma(\{i,j\}\times\{a,b\})} \sum_{\alpha=1}^{N_G} g_\alpha \Bigg[\Bigg(t^{A_\alpha}_{li}y_{kma}^{*} y_{mlb} 
    + \frac{1}{2}y_{lia}^{\phantom *} y_{kld}^* \theta^{A_\alpha}_{db}\nonumber\\ 
    &- \sum_{\beta=1}^{N_G} g_\beta^2 t^{B_\beta}_{ik} \theta^{B_\beta}_{da} \theta^{A_\alpha}_{db} \Bigg)\left[C_{\psi^2 F}\right]_{jk}^{A_\alpha}-\frac{1}{4} \theta_{ad}^{A_\alpha} y_{kib} y_{kld}^* \left[C_{\psi^2 F}\right]_{lj}^{A_\alpha}\Bigg]\,.\label{psi2phi2_psi2F}
\end{align}

\paragraph{\texorpdfstring{$\psi^2 \phi^2 \leftarrow \overline \psi^2 \psi^2$:}{psi2phi2frompsibar2psi2}}

\begin{align}
    \left[\dot C_{\psi^2\phi^2}\right]_{ijab} &= 4\sum_{\sigma(\{a,b\})} \left(m_{kn}y_{lma}y^*_{nlb}+y_{kna}m_{lm}y^*_{nlb}+y_{kna}y_{lmb} m^*_{nl}\right)\left[C_{\overline \psi^2 \psi^2}\right]_{kimj}
    \nonumber
    \\& -2 h_{abc}y_{klc} \left[C_{\overline \psi^2 \psi^2}\right]_{likj} \,.\label{psi2phi2_psibar2psi2}
\end{align}

\paragraph{\texorpdfstring{$\psi^2 \phi^2 \leftarrow \psi^4$:}{psi2phi2frompsi4}}

\begin{align}
    \left[\dot C_{\psi^2\phi^2}\right]_{ijab} &= -6 \sum_{\sigma(\{a,b\})}\left(m_{kn}y_{kma}^*y^*_{nlb}
    +y_{kna}m_{km}^*y^*_{nlb}+
    y_{kna}y_{kmb}^*m^*_{nl}\right)\left[C_{\psi^4}\right]_{ijlm}\nonumber\\
    &+ 3h_{abc}y^*_{klc} \left[C_{\psi^4}\right]_{ijkl}\,.
    \label{psi2phi2_psi4}
\end{align}

\paragraph{\texorpdfstring{$\psi^2 \phi^2 \leftarrow D^2\phi^4$:}{psi2phi2fromD2phi4}}

\begin{align}
    \left[\dot C_{\psi^2\phi^3}\right]_{ijab} &= \sum_{\sigma(\{a,b\})}
    y_{ike}y_{jld}m^*_{kl} \left[C_{D^2\phi^4}\right]_{deab}
    \nonumber \\&
    + \frac{1}{8} \sum_{\sigma(\{a,b\} \times \{i,j\})} y_{ike}m_{jl}y^*_{kld} \left(\left[C_{D^2\phi^4}\right]_{deab} + \left[C_{D^2\phi^4}\right]_{abde} 
    \right)\,.
    \label{psi2phi2_D2phi4}
\end{align}

\paragraph{\texorpdfstring{$\psi^2 \phi^2 \leftarrow D\overline\psi\psi\phi^2$}{psi2phi2fromDpsibarpsiphi2}:}

\begin{align}
    \left[\dot C_{\psi^2\phi^2}\right]_{ijab} &= 3\sum_{\sigma(\{i,j\}\times\{a,b\})}  \Bigg[\sum_{\alpha=1}^{N_G} g_\alpha^2 \left(\frac{1}{4} m_{ki} t_{lj}^{A_\alpha} \theta_{ad}^{A_\alpha} \left[C_{D\overline\psi\psi\phi^2}\right]_{klbd}\right.\nonumber\\
    &\left. - \frac{1}{3} m_{lk}t_{li}^{A_\alpha} \theta_{ad}^{A_\alpha} \left[C_{D\overline\psi\psi\phi^2}\right]_{kjbd}+ \frac{1}{12} m_{ki}t_{kl}^{A_\alpha} \theta_{ad}^{A_\alpha} \left[C_{D\overline\psi\psi\phi^2}\right]_{ljbd}\right)\nonumber\\
    &+ \frac{1}{3} m_{lk}^{\phantom *} \left( y_{mid}^{\phantom *} y_{lma}^* +\frac{1}{2}y_{mia}^{\phantom *} y_{lmd}^* \right)  \left[C_{D\overline\psi\psi\phi^2}\right]_{kjbd}\nonumber\\
    &+ \frac{1}{3} y_{lka}^{\phantom *} \left( y_{mid}^{\phantom *} m_{lm}^* +\frac{1}{2}m_{mi}^{\phantom *} y_{lmd}^* \right)  \left[C_{D\overline\psi\psi\phi^2}\right]_{kjbd}
    +\frac{1}{3} y_{kie} h_{ade} \left[C_{D\overline\psi\psi\phi^2}\right]_{kjbd}\Bigg]\label{psi2phi2_Dbarpsipsiphi2}\,.
\end{align}

\paragraph{\texorpdfstring{$\psi^2 \phi^2 \leftarrow \psi^2\phi^2\times\psi^2\phi^2$}{psi2phi2frompsi2phi2xpsi2phi2}:}

\begin{align}
    \left[\dot C_{\psi^2\phi^2}\right]_{ijab} &= -4\sum_{\sigma(\{i,j\}\times\{a,b\})}
    \bigg(
    m_{kl}^* \left[C_{\psi^2\phi^2}\right]_{ilad} \left[C_{\psi^2\phi^2}\right]_{jkbd}- 
    m_{jl} \left[C_{\psi^2\phi^2}\right]_{ikad} \left[C_{\psi^2\phi^2}\right]_{lkbd}^*
    \bigg)\,.\label{psi2phi2_psi2phi2xpsi2phi2}
\end{align}

\paragraph{$\psi^2\phi^2 \leftarrow \phi F^2 \times \phi F^2$:}

\begin{align}
    \left[\dot C_{\psi^2\phi^2}\right]_{ijab} &= \sum_{\sigma(\{i,j\}\times\{a,b\})} \sum_{\alpha=1}^{N_G} \sum_{\beta=1}^{N_G}\sum_{\gamma=1}^{N_G}\frac{3}{2} \mathcal{S}_{\alpha\gamma} \mathcal{S}_{\beta\gamma}g_\alpha g_\beta t_{ki}^{A_\alpha} \left[  i t_{lk}^{B_\beta} m_{lj} \left(\left[C_{\phi F^2}\right]_{a}^{B_\beta C_\gamma} \left[C_{\phi \tildee F^2}\right]_{b}^{A_\alpha C_\gamma}\right.\right.\nonumber\\
    &\left.\left.+ \left[C_{\phi F^2}\right]_{b}^{B_\beta C_\gamma} \left[C_{\phi \tildee F^2}\right]_{a}^{A_\alpha C_\gamma}\right) +  2 t_{lj}^{B_\beta} m_{kl} \left[C_{\phi F^2}\right]_a^{A_\alpha C_\gamma} \left[C_{\phi F^2}\right]_b^{B_\beta C_\gamma}\right]\,.
\end{align}

\paragraph{$\psi^2\phi^2 \leftarrow \psi^2 F \times \psi^2 F$:}

\begin{align}
    \left[\dot C_{\psi^2\phi^2}\right]_{ijab} &= 3\sum_{\sigma(\{i,j\}\times\{a,b\})} \Bigg[\sum_{\alpha=1}^{N_G} \sum_{\beta=1}^{N_G} 2 g_\alpha g_\beta \theta^{A_\alpha}_{ad}\theta^{B_\beta}_{db} m_{kl}^* \left[C_{\psi^2 F}\right]_{ki}^{A_\alpha}\left[C_{\psi^2 F}\right]_{lj}^{B_\beta}\nonumber\\
    & +\sum_{\alpha=1}^{N_G} \left(m_{kl}^{*} y_{lma}^{\phantom *} \, y^*_{mnb}+y_{kla}^{*} m_{lm}^{\phantom *} \, y^*_{mnb}\right) \left[C_{\psi^2 F}\right]_{ki}^{A_\alpha}\left[C_{\psi^2 F}\right]_{nj}^{A_\alpha}\Bigg]\,. 
\end{align}

\paragraph{$\psi^2\phi^2 \leftarrow \phi F^2 \times \psi^2 F$:}

\begin{align}
    \left[\dot C_{\psi^2\phi^2}\right]_{ijab} &= 3\sum_{\sigma(\{i,j\}\times\{a,b\})} \sum_{\alpha=1}^{N_G}\sum_{\beta=1}^{N_G} \mathcal{S}_{\alpha\beta} g_\alpha\Bigg[\frac{i}{2} m_{ki}^{\phantom *} t_{kl}^{A_\alpha} y_{lma}^* \left[C_{\phi \tildee F^2}\right]_{b}^{A_\alpha B_\beta}\left[C_{\psi^2 F}\right]_{mj}^{B_\beta}\nonumber\\
    &+ \frac{i}{2} y_{kia}^{\phantom *} t_{kl}^{A_\alpha} m_{lm}^* \left[C_{\phi \tildee F^2}\right]_{b}^{A_\alpha B_\beta}\left[C_{\psi^2 F}\right]_{mj}^{B_\beta}\nonumber\\
    &+ t_{ki}^{A_\alpha} \left(m_{kl}^{\phantom *} y_{lma}^* +y_{kla}^{\phantom *} m_{lm}^* \right)\left[C_{\phi F^2}\right]_{b}^{A_\alpha B_\beta}\left[C_{\psi^2 F}\right]_{mj}^{B_\beta}\Bigg]\,.
\end{align}

\paragraph{$\psi^2\phi^2 \leftarrow \psi^2\phi^2 \times \psi^2 F$:}

\begin{align}
    \left[\dot C_{\psi^2\phi}\right]_{ijab} &= 3\sum_{\sigma(\{i,j\}\times\{a,b\})} \sum_{\alpha=1}^{N_G} g_\alpha \theta_{ad}^{A_\alpha} m_{ki} \left[C_{\psi^2\phi^2}\right]_{klbd}^* \left[C_{\psi^2 F}\right]_{lj}^{A_\alpha}\nonumber \\
    &-3 
    \sum_{\sigma(\{i,j\})}\sum_{\alpha=1}^{N_G}
    g_\alpha \left(
    t^{A_\alpha}_{mk}m^*_{kl}\left[C_{\psi^2 F}\right]^{A_\alpha}_{li}+
    t^{A_\alpha}_{ki}m_{kl}\left[C_{\psi^2 F}\right]^{A_\alpha *}_{lm}\right)\left[C_{\psi^2\phi^2}\right]_{mjab}
    \,.
\end{align}

\subsubsection{\texorpdfstring{$\psi^2 F$ class}{psi2F class}}

\paragraph{$\psi^2 F \leftarrow \psi^2 \phi F$:}

\begin{align}
    \left[\dot C_{\psi^2 F}\right]_{ij}^{A_\alpha} &=  \left(2y_{jkb}m^*_{kl}+m_{jk}y^*_{klb}\right)\left[C_{\psi^2\phi F}\right]_{ilb}^{A_\alpha}\,.\label{psi2F_psi2F}
\end{align}

\paragraph{$\psi^2 F \leftarrow \psi^2 F$:}

\begin{align}
    \left[\dot C_{\psi^2 \phi F}\right]_{ij}^{A_\alpha} &=  \sum_{\beta=1}^{N_G} 8 g_\beta^2 t^{B_\beta}_{li}t^{B_\beta}_{kj} \left[ C_{\psi^2 F}\right]_{kl}^{A_\alpha} - y_{klb}^* y_{ikb}\left[ C_{\psi^2 F}\right]_{lj}^{A_\alpha} \nonumber \\
    &- \left(4\sum_{\beta=1}^{N_G}g_\alpha g_\beta t^{A_\alpha}_{kj}t^{B_\beta}_{lk}\left[C_{\psi^2\phi F}\right]^{B_\beta}_{il}
    -\left(i \leftrightarrow j\right)\right)\nonumber \\& + \gamma_{c,f}^{ik}\left[C_{\psi^2 F}\right]^{A_\alpha}_{kj} + \gamma_{c,f}^{jk}\left[C_{\psi^2 F}\right]^{A_\alpha}_{ik}    + \gamma_{c,v}^{A_\alpha B_\alpha}\left[C_{\psi^2 F}\right]^{B_\alpha}_{ij}
    \,.\label{psi2F_psi2phiF}
\end{align}
Here, $\gamma_{c,v}$ and $\gamma_{c,f}$ denote the collinear anomalous dimensions of gauge boson and fermion fields, respectively, as defined in Eqs.~\eqref{eq:coll_vector} and \eqref{eq:coll_fermion}.

\paragraph{$\psi^2  F \leftarrow \phi F^2$:}

\begin{equation}
    \left[\dot C_{\psi^2  F}\right]_{ij}^{A_\alpha} =
    \sum_{\beta=1}^{N_G} \mathcal S_{\alpha\beta} g_\beta \left(t^{B_\beta}_{ki} y_{jkb}-t^{B_\beta}_{kj} y_{ikb}  \right) \left(
    \left[C_{\phi^2 F^2}\right]_{b}^{A_\alpha B_\beta} + i \left[C_{\phi^2 \widetilde F^2}\right]_{b}^{A_\alpha B_\beta}
    \right)\,.\label{psi2F_phiF2}
\end{equation}

\paragraph{$\psi^2 F \leftarrow \psi^4$:}

\begin{equation}
    \left[\dot C_{\psi^2 F}\right]_{ij}^{A_\alpha} = - 4 g_\alpha \left(
    t^{A_\alpha}_{lm} m^*_{km} - t^{A_\alpha}_{km} m^*_{lm} 
    \right)
    \left[C_{\psi^4}\right]_{ikjl}
    \,.\label{psi2F_psi4}
\end{equation}

\paragraph{$\psi^2 F \leftarrow F^3$:}

\begin{align}
    \left[\dot C_{\psi^2  F}\right]_{ij}^{A_\alpha} &=  3ig_\alpha^2 \left(t^{C_\alpha}_{ki}t^{B_\alpha}_{mk}m_{jm} + 2 t^{C_\alpha}_{ki} t^{B_\alpha}_{mj}m_{km}
    - \left(i \leftrightarrow j \right)
    \right)\nonumber
    \\&\times 
    \left(
    \left[C_{F^3}\right]^{A_\alpha B_\alpha C_\alpha}+i \left[C_{\widetilde F^3}\right]^{A_\alpha B_\alpha C_\alpha}
    \right)\,.
    \label{psi2F_F3}
\end{align}

\paragraph{$\psi^2 F \leftarrow \phi F^2 \times \psi^2 F$:}

\begin{align}
    \left[\dot C_{\psi^2  F}\right]_{ij}^{A_\alpha} & =
    4 \sum_{\beta=1}^{N_G}\Bigg[ \left(y_{kib}^{\phantom *}m_{kl}^* + \frac{1}{2}m_{ki}^{\phantom *}y_{klb}^*\right)\left(\left[C_{\phi F^2}\right]_{b}^{A_\alpha B_\beta}+i\left[C_{\phi \tildee F^2}\right]_{b}^{A_\alpha B_\beta}\right)\left[C_{\psi^2 F}\right]_{lj}^{B_\beta}\nonumber\\
    &+ \frac{1}{2}y_{kib} m_{lj}\left(\left[C_{\phi F^2}\right]_{b}^{A_\alpha B_\beta}+i\left[C_{\phi \tildee F^2}\right]_{b}^{A_\alpha B_\beta}\right)\left[C_{\psi^2 F}\right]_{lk}^{B_\beta*}\Bigg] - (i\leftrightarrow j)
    \,.
\end{align}

\paragraph{$\psi^2 F \leftarrow \psi^2 F \times \psi^2 F$:}

\begin{align}
    \left[\dot C_{\psi^2 F}\right]_{ij}^{A_\alpha} & = 
    2i g_\alpha f^{A_\alpha B_\alpha C_\alpha}\left(
    2m^*_{km}\left[C_{\psi^2 F}\right]_{ki}^{B_\alpha}\left[C_{\psi^2 F}\right]_{mj}^{C_\alpha}
    + 3 m_{ik}\left[C_{\psi^2 F}\right]_{kl}^{B_\alpha *}\left[C_{\psi^2 F}\right]_{lj}^{C_\alpha}
    \right)
    \nonumber \\&
    +\sum_{\beta=1}^{N_G}\Bigg[
    2 g_\alpha t^{A_\alpha}_{kl} m^*_{lm}\left[C_{\psi^2 F}\right]_{ik}^{B_\beta}\left[C_{\psi^2 F}\right]_{mj}^{B_\beta}
    + 4 g_\alpha t^{A_\alpha}_{mj} m_{ik}\left[C_{\psi^2 F}\right]_{lm}^{B_\beta}\left[C_{\psi^2 F}\right]_{kl}^{B_\beta *}
    \nonumber \\&
    + g_\beta \Bigg(
    4 t^{B_\beta}_{mi} m^*_{kl}\left[C_{\psi^2 F}\right]_{lm}^{A_\alpha}\left[C_{\psi^2 F}\right]_{kj}^{B_\beta}
    -2 t^{B_\beta}_{ki} m_{mj}\left[C_{\psi^2 F}\right]_{kl}^{A_\alpha}\left[C_{\psi^2 F}\right]_{lm}^{B_\beta *}
    \nonumber \\&
    +6 t^{B_\beta}_{ml} m_{ik}\left[C_{\psi^2 F}\right]_{mj}^{A_\alpha}\left[C_{\psi^2 F}\right]_{kl}^{B_\beta *}
    + 6 t^{B_\beta}_{lk} m_{ik}\left[C_{\psi^2 F}\right]_{mj}^{A_\alpha}\left[C_{\psi^2 F}\right]_{lm}^{B_\beta *}
    \nonumber \\&
    -6 t^{B_\beta}_{ki}m_{kl}\left[C_{\psi^2 F}\right]_{mj}^{A_\alpha}\left[C_{\psi^2 F}\right]_{lm}^{B_\beta *} \Bigg)\Bigg] - \left(i \leftrightarrow j\right)\,.
\end{align}

\subsection{Running of renormalizable couplings}

\subsubsection{Gauge couplings and topological angles}

\paragraph{$F^2 \leftarrow \psi^2 F$:}

\begin{align}
    \dot g_\alpha \delta^{A_\alpha B_\alpha} &= 4\sum_{\sigma(\{A_\alpha,B_\alpha\})}
     g_\alpha^2 \Re{
     m^*_{ik}t^{B_\alpha}_{jk}
    \left[C_{\psi^2 F}\right]_{ij}^{A_\alpha}
    }
    \,,\label{g_psi2F}
    \\
    \dot \vartheta_\alpha \delta^{A_\alpha B_\alpha} &= \frac{32\pi^2}{g_\alpha^2} \,2\sum_{\sigma(\{A_\alpha,B_\alpha\})}
     g_\alpha \Im{
     m^*_{ik}t^{B_\alpha}_{jk}
    \left[C_{\psi^2 F}\right]_{ij}^{A_\alpha}
    }
    \,.\label{theta_psi2F}
\end{align}

\paragraph{$F^2\leftarrow \psi^2 F \times \psi^2 F$:}

\begin{align}
    \dot g_\alpha \delta^{A_\alpha B_\alpha} &= 16g_\alpha \Re{m^*_{ji} \left[C_{\psi^2 F}\right]_{il}^{A_\alpha} m^*_{lk} \left[C_{\psi^2 F}\right]_{kj}^{B_\alpha}} \,,\label{gauge_psi2Fxpsi2F}
    \\
    \dot \vartheta_\alpha \delta^{A_\alpha B_\alpha} &= \frac{32\pi^2}{g_\alpha^2}\, 8 \Im{m^*_{ji} \left[C_{\psi^2 F}\right]_{il}^{A_\alpha} m^*_{lk} \left[C_{\psi^2 F}\right]_{kj}^{B_\beta}} \,.\label{theta_psi2Fxpsi2F}
\end{align}

\subsubsection{Scalar quartic}

In addition to the results presented below, one needs to supplement the RGEs of the quartic scalar couplings with the expressions given in Eq.~\eqref{eq:phi6<-Ctildeandbar}.
The explicit expressions for $\dot{\tildee C}_{D^2\phi^4}$ and $\dot{\overline C}_{D^2\phi^4}$ can be obtained using their definitions in Eqs.~\eqref{eq:Ctilde_def} and \eqref{eq:Cbar_def} and the RGEs for $C_{D^2\phi^4}$ reported in Sec.~\ref{sec:D2phi4RGE}.

\paragraph{\texorpdfstring{$\phi^4\leftarrow \psi^2\phi^3$}{quarticfrompsi2phi3}:}

\begin{align}
    \dot \lambda_{abcd} &= 8\sum_{\sigma(\{a,b,c,d\})} \Re{\left(m_{ij}\, m_{jk}^*y_{kla}+\frac{1}{2}m_{ija} y_{jka}^* m_{kl}\right)\left[C_{\psi^2\phi^3}\right]_{libcd}^*}\,.\label{quartic_psi2phi3}
\end{align}

\paragraph{\texorpdfstring{$\phi^4\leftarrow \psi^2\phi^2$}{quarticfrompsi2phi2}:}

\begin{align}
    \dot \lambda_{abcd} &= 8\sum_{\sigma(\{a,b,c,d\})} \Re{\left(m_{ij}\, y_{jka}^*\, y_{klb}+y_{ija}\, m_{jk}^*\, y_{klb}\right)\left[C_{\psi^2\phi^2}\right]_{licd}^*}\nonumber\\
    & -\frac{1}{2}\sum_{\sigma(\{a,b,c,d\})} \Re{y_{ije}h_{abe}\left[C_{\psi^2\phi^2}\right]_{jicd}^*}\,.\label{quartic_psi2phi2}
\end{align}

\paragraph{\texorpdfstring{$\phi^4\leftarrow \psi^2\phi^2\times\psi^2\phi^2$}{quarticfrompsi2phi2xpsi2phi2}:}

\begin{align}
    \dot \lambda_{abcd} &= -8 \sum_{\sigma(\{a,b,c,d\})}\Bigg[\Re{\left[C_{\psi^2\phi^2}\right]_{ijab} m^*_{jk} \left[C_{\psi^2\phi^2}\right]_{klcd} m^*_{li}}
    \nonumber\\
     &
     +2 m^*_{jk} m_{kl} \left[C_{\psi^2\phi^2}\right]_{ijab}\left[C_{\psi^2\phi^2}\right]^*_{licd}\Bigg]\,.\label{quartic_psi2phi2xpsi2phi2}
\end{align}

\subsubsection{Scalar trilinear}

In addition to the results presented below, one needs to supplement the RGEs of the trilinear scalar couplings with the expressions given in Eq.~\eqref{eq:phi6<-Ctildeandbar}.
The explicit expressions for $\dot{\tildee C}_{D^2\phi^4}$ and $\dot{\overline C}_{D^2\phi^4}$ can be obtained using their definitions in Eqs.~\eqref{eq:Ctilde_def} and \eqref{eq:Cbar_def} and the RGEs for $C_{D^2\phi^4}$ reported in Sec.~\ref{sec:D2phi4RGE}.

\paragraph{\texorpdfstring{$\phi^3\leftarrow \psi^2\phi^3$}{trilinearfrompsi2phi3}:}

\begin{align}
    \dot h_{abc} &= 24\Re{m_{ij}\, m_{jk}^*m_{kl}\left[C_{\psi^2\phi^3}\right]_{liabc}^*}\,.\label{trilinear_psi2phi3}
\end{align}

\paragraph{\texorpdfstring{$\phi^3\leftarrow \psi^2\phi^2$}{trilinearfrompsi2phi2}:}

\begin{align}
    \dot h_{abc} &= 8\sum_{\sigma(\{a,b,c\})} \Re{\left(m_{ij}\, m_{jk}^*\, y_{kla}+\frac{1}{2}m_{ij}\, y_{jka}^*\, m_{kl}+y_{ija}m_{jk}^* m_{kl}\right)\left[C_{\psi^2\phi^2}\right]_{liab}^*}\nonumber\\
    & -\sum_{\sigma(\{a,b,c\})} \Re{y_{ijd}m^2_{ad}\left[C_{\psi^2\phi^2}\right]_{jibc}^*}\,.\label{trilinear_psi2phi2}
\end{align}

\subsubsection{Scalar mass}

\paragraph{\texorpdfstring{$\phi^2\leftarrow \psi^2\phi^2$}{smassfrompsi2phi2}:}

\begin{align}
    \dot{(m^2)}_{ab} &= 8\Re{m_{ij}\, m_{jk}^*m_{kl}\left[C_{\psi^2\phi^2}\right]_{liab}^* -\frac{1}{4} y_{ijc}t_{c}\left[C_{\psi^2\phi^2}\right]_{jiab}^*}\,.\label{smass_psi2phi3}
\end{align}



\subsubsection{Yukawa coupling}

\paragraph{\texorpdfstring{$\psi^2 \phi \leftarrow \psi^2 \phi^3$}{yukfrompsi2phi3}:}

\begin{align}
    \dot y_{ija} &= -12m^2_{bc}\left[ C_{\psi^2\phi^3}\right]_{ijabc}\,.\label{yuk_psi2phi3}
\end{align}

\paragraph{\texorpdfstring{$\psi^2 \phi \leftarrow \psi^2 \phi^2$}{yukfrompsi2phi2}:}

\begin{align}
    \dot y_{ija} &= -4h_{abc}\left[ C_{\psi^2\phi^2}\right]_{ijbc} -4\sum_{\sigma(\{i,j\})}\left(2y_{jkb}m^*_{kl}+m_{jk}y^*_{klb}\right)\left[ C_{\psi^2\phi^2}\right]_{ilab} \nonumber \\
    &-4\sum_{\sigma(\{i,j\})} m_{jl} y_{ikb} \left[C_{\psi^2\phi^2}\right]_{lkab}\,.\label{yuk_psi2phi2}
\end{align}

\paragraph{\texorpdfstring{$\psi^2 \phi \leftarrow \phi F^2$:}{yukfromphiF2}}

\begin{align}
    \dot y_{ija} &= -12\sum_{\sigma(\{i,j\})} \sum_{\alpha=1}^{N_G}\sum_{\beta=1}^{\alpha} g_\alpha g_\beta\Bigg[ t^{A_\alpha}_{ki} t^{B_\beta}_{lk} m_{lj} \left(\left[C_{\phi F^2}\right]_{a}^{A_\alpha B_\beta}+i\left[C_{\phi \tildee F^2}\right]_{a}^{A_\alpha B_\beta} \right)\nonumber\\
    &+t_{ki}^{A_\alpha} t_{lj}^{B_\beta} m_{kl}\left[C_{\phi F^2}\right]^{A_\alpha B_\beta}_a\Bigg]\,.
    \label{yuk_phiF2}
\end{align}

\paragraph{\texorpdfstring{$\psi^2 \phi \leftarrow \psi^2 \phi F$:}{yukfromphi2phiF}}

\begin{align}
    \dot y_{ija} &= -12\sum_{\sigma(\{i,j\})} \sum_{\alpha=1}^{N_G} g_\alpha t^{A_\alpha}_{il}m_{km}^{\phantom *} m^*_{ml} \left[C_{\psi^2\phi F}\right]_{jka}^{A_\alpha}\,.\label{yuk_psi2phiF}
\end{align}

\paragraph{\texorpdfstring{$\psi^2 \phi \leftarrow \psi^2 F$:}{yukfromphi2F}}

\begin{align}
    \dot y_{ija} &= -12\sum_{\sigma(\{i,j\})} \sum_{\alpha=1}^{N_G} g_\alpha \Bigg[\Bigg(t^{A_\alpha}_{li}(m_{km}^{*} y_{mla}+y_{kma}^{*} m_{ml}) 
    + \frac{1}{2}m_{li}^{\phantom *} y_{kld}^* \theta^{A_\alpha}_{da}\Bigg)\left[C_{\psi^2F}\right]_{jk}^{A_{\alpha}}\nonumber\\ 
    &-\frac{1}{4} \theta_{ad}^{A_\alpha} m_{ki} y_{kld}^* \left[C_{\psi^2 F}\right]_{lj}^{A_\alpha}
    +\frac{1}{4}y_{mja} \left(
    t^{A_\alpha}_{mk}m^*_{kl}\left[C_{\psi^2 F}\right]^{A_\alpha}_{li}+
    t^{A_\alpha}_{ki}m_{kl}\left[C_{\psi^2 F}\right]^{A_\alpha *}_{lm}\right)
    \Bigg]
    \,.\label{yuk_psi2F}
\end{align}

\paragraph{\texorpdfstring{$\psi^2 \phi \leftarrow \overline \psi^2 \psi^2$:}{yukfrompsibar2psi2}}

\begin{align}
    \dot y_{ija} &= -16\left(\frac{1}{2} m_{kn}m_{lm}y^*_{nla}+m_{kn}y_{lma} m^*_{nl}\right)\left[C_{\overline \psi^2 \psi^2}\right]_{kimj}
    \nonumber\\
    &
    +8 m_{ab}^2 y_{klb} \left[C_{\overline \psi^2 \psi^2}\right]_{likj} \,.\label{yuk_psibar2psi2}
\end{align}

\paragraph{\texorpdfstring{$\psi^2 \phi \leftarrow \psi^4$:}{yukfrompsi4}}

\begin{align}
   \dot y_{ija} &= 24 \left(m_{kn}m_{km}^*y^*_{nla}+
    \frac{1}{2}y_{kna}m_{km}^*m^*_{nl}\right)\left[C_{\psi^4}\right]_{ijlm}
    - 12m^2_{ab}y^*_{klb} \left[C_{\psi^4}\right]_{ijkl}\,.
    \label{yuk_psi4}
\end{align}

\paragraph{$\psi^2 \phi \leftarrow D^2\phi^4$:}

\begin{align}
    \dot y_{ija} = 4 y_{ijb} m^2_{cd} \left(\left[C_{D^2\phi^4}\right]_{cbad} - \left[C_{D^2\phi^4}\right]_{cdab} \right) \,.
\end{align}

\paragraph{\texorpdfstring{$\psi^2 \phi \leftarrow D\overline\psi\psi\phi^2$}{yukfromDpsibarpsiphi2}:}

\begin{align}
    \dot y_{ija} &= -12\sum_{\sigma(\{i,j\})}  \Bigg[\frac{1}{3} m_{lk}^{\phantom *} \left( y_{mid}^{\phantom *} m_{lm}^* +\frac{1}{2}m_{mi}^{\phantom *} y_{lmd}^* \right)  \left[C_{D\overline\psi\psi\phi^2}\right]_{kjad}
    \nonumber\\& 
    +\frac{1}{3} y_{kie} m^2_{de} \left[C_{D\overline\psi\psi\phi^2}\right]_{kjad}\Bigg]\label{yuk_Dbarpsipsiphi2}\,.
\end{align}

\paragraph{$\psi^2\phi \leftarrow \psi^2 F \times \psi^2 F$:}

\begin{align}
    \dot y_{ija} &= -12\sum_{\sigma(\{i,j\})} \sum_{\alpha=1}^{N_G} m_{kl}^{*} \Big(m_{lm}^{\phantom *} y^*_{mna}+y_{lma}^{\phantom *} m^*_{mn}\Big) \left[C_{\psi^2 F}\right]_{ki}^{A_\alpha}\left[C_{\psi^2 F}\right]_{nj}^{A_\alpha} \nonumber \\
    &-12 \sum_{\sigma(\{i,j\})} \sum_{\alpha=1}^{N_G} y_{kja} m_{nl}m^*_{lm}\left[C_{\psi^2 F}\right]^{A_\alpha}_{mi}\left[C_{\psi^2 F}\right]^{A_\alpha *}_{kn} \,. 
\end{align}

\paragraph{$\psi^2\phi \leftarrow \phi F^2 \times \psi^2 F$:}

\begin{align}
    \dot y_{ija} &= -12\sum_{\sigma(\{i,j\})} \sum_{\alpha=1}^{N_G}\sum_{\beta=1}^{N_G} \mathcal{S}_{\alpha\beta} g_\alpha\Bigg[\frac{i}{2} m_{ki}^{\phantom *} t_{kl}^{A_\alpha} m_{lm}^* \left[C_{\phi \tildee F^2}\right]_{a}^{A_\alpha B_\beta}\left[C_{\psi^2 F}\right]_{mj}^{B_\beta}\nonumber\\
    &+ \frac{i}{2} m_{ki}^{\phantom *} t_{kl}^{A_\alpha} m_{lm}^* \left[C_{\phi \tildee F^2}\right]_{a}^{A_\alpha B_\beta}\left[C_{\psi^2 F}\right]_{mj}^{B_\beta}\nonumber\\
    &+ t_{ki}^{A_\alpha} \left(m_{kl}^{\phantom *} y_{lm}^* +y_{kl}^{\phantom *} m_{lm}^* \right)\left[C_{\phi F^2}\right]_{a}^{A_\alpha B_\beta}\left[C_{\psi^2 F}\right]_{mj}^{B_\beta}\Bigg]\,.
\end{align}

\subsubsection{Fermion mass}

\paragraph{\texorpdfstring{$\psi^2 \leftarrow \psi^2 \phi^2$}{fmassfrompsi2phi2}:}

\begin{align}
    \dot m_{ij} &= -4m_{ab}^2\left[ C_{\psi^2\phi^2}\right]_{ijab} \,.\label{fmass_psi2phi2}
\end{align}

\paragraph{\texorpdfstring{$\psi^2 \leftarrow \psi^2 F$:}{fmassfromphi2F}}

\begin{align}
    \dot m_{ij} &= -12\sum_{\sigma(\{i,j\})} \sum_{\alpha=1}^{N_G} g_\alpha 
    \Bigg[t^{A_\alpha}_{il}m_{km}^{\phantom *} m^*_{ml}\left[C_{\psi^2F}\right]_{jk}^{A_{\alpha}}
    \nonumber \\&+\frac{1}{4}m_{mj} \left(
    t^{A_\alpha}_{mk}m^*_{kl}\left[C_{\psi^2 F}\right]^{A_\alpha}_{li}+
    t^{A_\alpha}_{ki}m_{kl}\left[C_{\psi^2 F}\right]^{A_\alpha *}_{lm}\right)
    \Bigg]\,.\label{fmass_psi2F}
\end{align}

\paragraph{\texorpdfstring{$\psi^2 \leftarrow \overline \psi^2 \psi^2$:}{fmassfrompsibar2psi2}}

\begin{align}
    \dot m_{ij} &= -8m_{kn}m_{lm}m^*_{nl}\left[C_{\overline \psi^2 \psi^2}\right]_{kimj}
    +8 t_a y_{kla} \left[C_{\overline \psi^2 \psi^2}\right]_{likj} \,.\label{fmass_psibar2psi2}
\end{align}

\paragraph{\texorpdfstring{$\psi^2 \leftarrow \psi^4$:}{fmassfrompsi4}}

\begin{align}
   \dot m_{ij} &= 12 m_{kn}m_{km}^*m^*_{nl}\left[C_{\psi^4}\right]_{ijlm}- 12 t_{a}y^*_{kla} \left[C_{\psi^4}\right]_{ijkl}\,.
    \label{fmass_psi4}
\end{align}

\paragraph{\texorpdfstring{$\psi^2 \leftarrow \psi^2 F \times \psi^2 F$}{fmassfrompsi2F}:}

\begin{align}
    \dot m_{ij} &= -6\sum_{\sigma(\{i,j\})} \sum_{\alpha=1}^{N_G} m_{kl}^{*} m_{lm}^{\phantom *} m^*_{mn} \left[C_{\psi^2 F}\right]_{ki}^{A_\alpha}\left[C_{\psi^2 F}\right]_{nj}^{A_\alpha} \nonumber \\
    &-12 \sum_{\sigma(\{i,j\})} \sum_{\alpha=1}^{N_G} m_{kj} m_{nl}m^*_{lm}\left[C_{\psi^2 F}\right]^{A_\alpha}_{mi}\left[C_{\psi^2 F}\right]^{A_\alpha *}_{kn}\,. 
\end{align}

\section{Conclusions and outlook}\label{sec:concl}
In this article, we presented the complete set of RGEs for arbitrary EFTs, including operators with fermionic fields up to mass dimension six. For this purpose, a general setup including an arbitrary number of scalar, fermion, and vector fields transforming under arbitrary gauge groups was adopted in order to obtain the corresponding template RGEs. We reported the complete one-loop results for all renormalizable and non-renormalizable couplings up to mass dimension six, where RGE effects of $\mathcal{O}(1/\Lambda^2)$ in the EFT power counting were consistently taken into account. Together with the bosonic RGEs obtained in \cite{Aebischer:2025zxg}, these results conclude the derivation of the complete set of one-loop RGEs in arbitrary EFTs at the dimension-six level. Consequently, this article provides the framework for deriving the one-loop RGEs of any conceivable EFT at the one-loop level. 

This opens a possibility to study new physics scenarios in which the Standard Model is extended by new light degrees of freedom that can imprint distinctive signatures via Renormalization Group mixing into the well-constrained SMEFT operators~\cite{Galda:2021hbr}. Therefore, automating the matching of the results derived here to the specific EFTs would be of immediate use, offering a systematic way to probe light new states largely independently of their mass or lifetime. As a next step, we envisage such an implementation interfaced with dedicated RGE solvers such as \texttt{wilson} \cite{Aebischer:2018bkb} or \texttt{DSixtools} \cite{Celis:2017hod,Fuentes-Martin:2020zaz}, enabling straightforward phenomenological analyses.

Moreover, the majority of the 184 operator mixings identified (131 derived here and 53 in the bosonic sector in Ref.~\cite{Aebischer:2018bkb}) exhibit a non-trivial structure that becomes transparent within the general EFT framework employed in this work. A detailed analysis of the observed patterns, the emergence of new selection rules, and the holomorphic structure of the most general anomalous dimension matrix is deferred to separate work. Finally, the results obtained here, as well as their extensions to higher orders, may have interesting implications for conformal field theories, where RGEs can be used to extract the scaling dimensions of composite operators \cite{Henriksson:2025hwi}.

As a further step, the two-loop RGEs for general EFTs can be derived in a similar manner, as was performed, for instance, for the bosonic sector in \cite{Guedes:2025sax}. However, in the case of fermionic operators, issues regarding evanescent operators \cite{Aebischer:2022tvz,Aebischer:2022aze,Aebischer:2022rxf} and renormalization schemes \cite{Aebischer:2023djt,Aebischer:2024xnf} need to be taken into account. We leave this work for the future.

\acknowledgments 
We thank Barbara Anna Erdelyi for useful checks on some of the results.
LCB and NS thank Paride Paradisi for helpful discussions and encouragement.
The work of J.A. was funded by the Swiss National Science Foundation (SNSF) through grant~TMSGI2-225951.
NS is supported by the Italian MUR through the FIS 2 project FIS-2023-01577 (DD n. 23314 10-12-2024, CUP C53C24001460001), and by Istituto Nazionale di Fisica Nucleare (INFN) through the Theoretical Astroparticle Physics (TAsP) project.
The project received funding from the INFN Iniziativa Specifica APINE.

\appendix
\newpage

\section{Conventions and definitions}\label{app:conventions}

In this appendix the relevant conventions and definitions used in this work are reported.

\subsection{Collinear anomalous dimensions}\label{sec:coll_ADM}

The infrared collinear anomalous dimensions, relevant for the multiplicative renormalization factors, are computed by using the on-shell method reviewed in Sec.~\ref{sec:method} and exploiting the fact that the stress-energy tensor does not renormalize multiplicatively, as done in \cite{Caron-Huot:2016cwu,EliasMiro:2020tdv,AccettulliHuber:2021uoa,Bresciani:2024shu,Baratella:2022nog,Aebischer:2025zxg}.
They are given by
\begin{align}
    \gamma_{c,v}^{A_\alpha B_\alpha} &= -g_\alpha^2 \left[b_{0,\alpha}\right]^{A_\alpha B_\alpha} \,, \label{eq:coll_vector}  \\
    \gamma_{c,s}^{ab} &= -4\sum_{\alpha=1}^{N_G} g_\alpha^2 \left[C_2(S_\alpha)\right]_{a_{\alpha}b_{\alpha}} + \frac{1}{2}\left(y_{ija}y_{jib}^*+y_{ijb}y_{jia}^*\right) \,, \label{eq:coll_scalar} \\
    \gamma_{c,f}^{ij} &= -3\sum_{\alpha=1}^{N_G} [C_2(F_\alpha)]_{i_\alpha j_\alpha}+\frac{1}{2}y_{ika}y^*_{kja} \,, \label{eq:coll_fermion}
\end{align}
for gauge boson, scalar, and fermion fields, respectively,
with the one-loop $\beta$-function coefficient
\begin{equation}
    \left[b_{0,\alpha}\right]^{A_\alpha B_\alpha} = \frac{11}{3}\left[C_2(G_\alpha)\right]^{A_\alpha B_\alpha} - \frac{1}{6} \left[S_2(S_\alpha)\right]^{A_\alpha B_\alpha} -\frac{2}{3}[S_2(F_\alpha)]^{A_\alpha B_\alpha}\,.
\end{equation}
The quadratic Casimir $C_2$ and Dynkin indices $S_2$ for the different representations are given by
\begin{align}
    \left[C_2(G_\alpha)\right]^{A_\alpha B_\alpha} &= f^{A_\alpha C_\alpha D_\alpha}f^{B_\alpha C_\alpha D_\alpha}\,,\\
    \left[S_2(S_\alpha)\right]^{A_\alpha B_\alpha} &= \Tr(\theta^{A_\alpha}\theta^{B_\alpha})\,, \\ 
    \left[C_2(S_\alpha)\right]_{a_{\alpha}b_{\alpha}} &= \theta^{A_\alpha}_{a_{\alpha}c_{\alpha}} \theta^{A_\alpha}_{c_{\alpha}b_{\alpha}}\,,\\
    \left[S_2(F_\alpha)\right]^{A_\alpha B_\alpha} &= \Tr(t^{A_\alpha}t^{B_\alpha})\,, \\
    \left[C_2(F_\alpha)\right]_{i_{\alpha}j_{\alpha}} &= t^{A_\alpha}_{j_{\alpha}k_{\alpha}} t^{A_\alpha}_{k_{\alpha}i_{\alpha}}\,.
\end{align}

\subsection{Symmetry factors}
\label{sec:symm_factors}
To account for symmetry factors, the following shorthand notation is introduced:
\begin{equation}
\mathcal S_{i_1\dots i_n}=\prod_{\ell=1}^{N_{i_1\dots i_n}} k_\ell !\,,
\label{eq:symm_fact}
\end{equation}
where $I=\{i_1,\dotsc, i_n\}$ denotes the multiset of indices.
The quantity $N_{i_1\dots i_n}$ is the number of distinct elements in $I$, and $k_\ell$ is the multiplicity of the $\ell$-th distinct index.
By construction
\begin{equation}
\sum_{\ell=1}^{N_{i_1\dots i_n}}k_\ell=n\,.
\end{equation}
As an example, one finds for $n=2$ and $n=3$
\begin{align}
\mathcal S_{ij} = \begin{cases}
1 & \text{if }i\neq j, \\
2 & \text{if }i=j,
\end{cases} && 
\mathcal S_{ijk} =
\begin{cases}
1 & \text{for } i,j,k \text{ all distinct}, \\
2 & \text{for exactly two equal indices}, \\
6 & \text{for } i=j=k.
\end{cases}
\end{align}

\bibliographystyle{JHEP}
\bibliography{biblio}

\end{document}